\documentclass[aps,prd,superscriptaddress]{revtex4}
\usepackage{epsfig,epsf}
\usepackage{amsmath}
\usepackage{amsthm}
\usepackage{amsfonts}
\usepackage{amssymb}
\usepackage{dsfont}
\usepackage{multirow}
\usepackage{appendix}
\usepackage{slashed}
\usepackage[active]{srcltx}
\usepackage{psfrag}
\usepackage{multirow}
\def\beq{\begin{equation}}
\def\eeq{\end{equation}}
\def\bea{\begin{eqnarray}}
\def\eea{\end{eqnarray}}
\def\beeq{\begin{eqnarray}}
\def\eeeq{\end{eqnarray}}

\def\nnb{\nonumber}

\def\rar{\rightarrow}
  
\def\nnb{\nonumber}

\def\ba{\begin{array}}
\def\ea{\end{array}}

\def\xis0{{\Xi^{*0}}}

\def\g5{\gamma_5}

\def\baeq{\begin{appeq}}     \def\eaeq{\end{appeq}}  
\def\baeeq{\begin{appeeq}}   \def\eaeeq{\end{appeeq}}
\newenvironment{appeq}{\beq}{\eeq}   
\newenvironment{appeeq}{\beeq}{\eeeq}

\def\eAPP{\renewcommand{\thehran}{\thesection.\arabic{hran}}}

\renewcommand{\theequation}{\arabic{equation}}
\newcounter{hran}
\renewcommand{\thehran}{\thesection.\arabic{hran}}

\def\bmini{\setcounter{hran}{\value{equation}}
\refstepcounter{hran}\setcounter{equation}{0}
\renewcommand{\theequation}{\thehran\alph{equation}}\begin{eqnarray}}
\def\bminiG#1{\setcounter{hran}{\value{equation}}
\refstepcounter{hran}\setcounter{equation}{-1}
\renewcommand{\theequation}{\thehran\alph{equation}}
\refstepcounter{equation}\label{#1}\begin{eqnarray}}

\newskip\humongous \humongous=0pt plus 1000pt minus 1000pt

\begin{document}

\title{Determination of the quantum numbers of $\Sigma_b(6097)^{\pm}$ via their strong decays}
\date{\today}
\author{T.~M.~Aliev}
\affiliation{Physics Department,
Middle East Technical University, 06531 Ankara, Turkey}
\author{K.~Azizi}
\affiliation{Physics Department, Do\u gu\c s University,
Ac{\i}badem-Kad{\i}k\"oy, 34722 Istanbul, Turkey}
\author{Y.~Sarac}
\affiliation{Electrical and Electronics Engineering Department,
Atilim University, 06836 Ankara, Turkey}
\author{H.~Sundu}
\affiliation{Department of Physics, Kocaeli University, 41380 Izmit, Turkey}

\begin{abstract}
The progresses in the experimental sector have been the harbinger of the observations of many new hadrons. Very recently, LHCb Collaboration announced the observation of two new $\Sigma_b(6097)^{\pm}$ states in the  $\Lambda^0_b\pi^{\pm}$ invariant mass distribution, which are considered as the excited states of the ground state $\Sigma^{(*)}_b$ baryon. Though, almost all of the ground state baryons have been observed, having a limited number of excited states observed so far makes them intriguing. Understanding the properties of the excited baryons improve our knowledge on the strong interaction as well as the  nature and internal structures of these baryons. To specify the  quantum numbers of the $\Sigma_b(6097)^{\pm}$ an analysis on their strong decays to $\Lambda_b^0$ and $\pi^{\pm}$ is performed within the light cone QCD sum rule formalism. To this end, they are considered as possible $1P$ or $2S$ excitation of either the ground state $\Sigma_b$ baryon with $J=\frac{1}{2}$ or $\Sigma_b^{*}$ baryon with $J=\frac{3}{2}$. The corresponding masses are also calculated considering the same  scenarios for their quantum numbers. The results of the analyses indicate that the $\Sigma_b(6097)^{\pm}$ baryons are excited $1P$ baryons having quantum numbers $J^P=\frac{3}{2}^-$.

\end{abstract}

\maketitle
\section{Introduction}
In the quark model, the heavy baryons containing one heavy and two light quarks form multiplets using the symmetry of flavor, spin, and spatial wave functions~\cite{Klempt:2009pi}. These considerations lead to the results that they belong to the sextet and the antitriplet representations of $SU(3)$. At present, almost all the ground-state heavy baryons have been observed in experiments. According to the quark model predictions, in addition to the ground states, the existence of their excited states is also expected. So far, only a few excited baryons have been observed in the bottom sector~\cite{Aaij:2015yoy,Aaij:2018yqz,Aaij:2012da,Chatrchyan:2012ni,Aaij:2014yka}.
The detailed study of the experimentally discovered states and looking for new, yet to be observed, states can play a critical role for understanding of the internal structures of these states and give essential information about the dynamics of QCD at the non-perturbative domain. 

Very recently, the LHCb Collaboration has announced the first observation of two $\Sigma_b(6097)^-$ and $\Sigma_b(6097)^+$ resonances with masses $m(\Sigma_b(6097)^-)=6098.0\pm1.7\pm0.5$~MeV and $m(\Sigma_b(6097)^+)=6095.8\pm1.7\pm0.4$~MeV~\cite{Aaij:2018tnn}. The widths of these states have also been measured as $\Gamma(\Sigma_b(6097)^-)=28.9\pm4.2\pm0.9$~MeV  and $\Gamma(\Sigma_b(6097)^+)=31.0\pm5.5\pm0.7$~MeV. After the discovery of these states, the determination of their quantum numbers stands as a central problem. In Ref.~\cite{Chen:2018vuc}, to understand the structure of $\Sigma_b(6097)$, the mass and strong decay analyses were considered within a quasi-two-body treatment. As a result of this study, the $\Sigma_b(6097)$ was concluded to be a bottom baryon candidate having $J^P=\frac{3}{2}^-$ or $J^P=\frac{5}{2}^-$. In another study, the constituent quark model was applied to investigate $\Sigma_b(6097)$. The authors concluded that this state is a $P$-wave baryon with the quantum numbers $J^P=\frac{3}{2}^-$ or $J^P=\frac{5}{2}^-$~\cite{Wang:2018fjm}. Another prediction for the quantum numbers of the observed $\Sigma_b(6097)^-$ and $\Sigma_b(6097)^+$ states was presented in Ref.~\cite{Yang:2018lzg} via the quark-pair creation model, which indicated the possibility of their being again either $J^P=\frac{3}{2}^-$ or $J^P=\frac{5}{2}^-$.    

In the present study, the properties of these baryons are studied in the framework of the QCD sum rule method~\cite{Shifman:1978bx}. In our calculations, the observed states are considered as $1P$ or $2S$ excitated states with $J=\frac{1}{2}$ or $J=\frac{3}{2}$. We analyze the $\Sigma_b^{\pm}\rightarrow \Lambda_b \pi^{\pm}$ decays and compare the values of the obtained decay widths with the experimental results, which allows us to determine the quantum numbers of $\Sigma_b(6097)^{\pm}$ states. To calculate the decay widths the main ingredient is the coupling constants corresponding the considered transitions. For calculation of these coupling constants we use the light cone QCD sum rules (LCSR) method~\cite{Braun:1988qv}. In this work, we also calculate the masses and the decay constants of the states under consideration by taking into account again all possibilities, i.e. assuming that these states are $1P$ or $2S$ excited states of the ground state $\Sigma_b$ and  $\Sigma_b^{*}$ baryons with $J=\frac{1}{2}$ or $J=\frac{3}{2}$. The obtained  masses and  decay constants are used as inputs in the numerical computations of the strong coupling constants of the related decays. Similar coupling constants  for the ground state baryons with single heavy quark having $J=\frac{1}{2}$ and $J=\frac{3}{2}$  have been calculated in Refs.~\cite{Aliev:2010yx,Aliev:2010ev,Aliev:2011ufa,Azizi:2008ui}.  

The paper is organized as follows: In Section 2, the strong decays $\Sigma_b(6097)^{\pm}\rightarrow \Lambda_b^{0} \pi^{\pm}$ are studied within the LCSR method~\cite{Braun:1988qv} by taking into account the possible configurations assigned to the $\Sigma_b(6097)^{\pm}$ states. In this section, we also formulate the sum rules for the masses and  decay constants of $\Sigma_b(6097)^{\pm}$ with  $J=\frac{1}{2}$ or $J=\frac{3}{2}$. The numerical results of the  masses and decay constants  are  used as input parameters in the analyses of the strong coupling constants defining the above strong decay channels. The numerical results of the strong coupling constants are also used to obtain the numerical values of the decay widths of the transitions under consideration. The last section contains our concluding remarks.  The details of the calculations of the spectral densities are given in the appendix.

\section{Analysis of  the $\Sigma_b\Lambda_b \pi$ vertex via light cone QCD sum rule}

 In this section, we analyze the strong transitions of the $\Sigma_b(6097)^{\pm}$ states to the $\Lambda_b^0$ and $\pi^{\pm}$ particles. As we have already noted, our primary goal is to determine  the quantum numbers of the recently observed $\Sigma_b(6097)^{\pm}$ baryons. To this end, we assume that these states are $1P$ or $2S$ excitations of the  the corresponding ground-state baryons with $J=\frac{1}{2}$ or  $J=\frac{3}{2}$. We calculate the  widths of  these baryons under these assumptions and compare our results with that of the experimental data. 
 
 Each decay is characterized by its own strong coupling constant. Therefore, in the first step, we calculate the  corresponding coupling constant defining the strong $\Sigma_b\rightarrow \Lambda_b \pi$ transition for each case within the LCSR.  For the ground-state $\Sigma_b$ and $\Sigma_b^{*}$ particles, these strong coupling constants are defined  as
\begin{eqnarray}
\langle \pi(q)\Lambda_b(p,s)|\Sigma_b(p^{\prime },s^{\prime })\rangle &=& g_{\Sigma_b\Lambda_b \pi}\bar{u}(p,s) \gamma_5 u(p',s'),\nonumber\\
\langle \pi(q)\Lambda_b(p,s)|\Sigma_b^{*}(p^{\prime },s^{\prime })\rangle &=& g_{\Sigma_b^{*}\Lambda_b\pi}\bar{u}(p,s)u_{\mu}(p',s')q^{\mu}.\label{eq:Matrixelements1}
\end{eqnarray}
For their corresponding $1P$ and $2S$ excitations,  similar definitions as Eq.~(\ref{eq:Matrixelements1}) with the following replacements are used: 

$a)$ For the $1P$ excitations: $g_{\Sigma_b\Lambda_b \pi}\rightarrow g_{\Sigma_{b1}\Lambda_b \pi}$, $g_{\Sigma_b^*\Lambda_b \pi}\rightarrow g_{\Sigma^*_{b1}\Lambda_b \pi}$, $ u(p',s')\rightarrow \gamma_5 u(p',s')$, $u_{\mu}(p',s')\rightarrow \gamma_5 u_{\mu}(p',s')$, $|\Sigma_b(p^{\prime },s^{\prime })\rangle \rightarrow |\Sigma_{b1}(p^{\prime },s^{\prime })\rangle $ and 
$|\Sigma_b^{*}(p^{\prime },s^{\prime })\rangle \rightarrow |\Sigma_{b1}^{*}(p^{\prime },s^{\prime })\rangle$,

$b)$ For the $2S$ excitations : $g_{\Sigma_b\Lambda_b \pi}\rightarrow g_{\Sigma_{b2}\Lambda_b \pi}$, $g_{\Sigma_b^*\Lambda_b \pi}\rightarrow g_{\Sigma^*_{b2}\Lambda_b \pi}$, $|\Sigma_b(p^{\prime },s^{\prime })\rangle \rightarrow |\Sigma_{b2}(p^{\prime },s^{\prime })\rangle $ and 
$|\Sigma_b^{*}(p^{\prime },s^{\prime })\rangle \rightarrow |\Sigma_{b2}^{*}(p^{\prime },s^{\prime })\rangle$.

In this section and in all the following discussions, the ground state and its  $1P$ and $2S$  excitations are denoted by $\Sigma_b(\Sigma_b^{*})$, $\Sigma_{b1}(\Sigma_{b1}^{*})$ and $\Sigma_{b2}(\Sigma_{b2}^{*})$ for corresponding $J=\frac{1}{2}(\frac{3}{2})$ baryons, respectively.  Here,  $u(q,s)$ and $ u_{\mu}(q,s)$ are  spinors corresponding to the $J=\frac{1}{2}$ and  $J=\frac{3}{2}$ states, respectively. 

For the determination of the aforementioned coupling constants from the LCSR, we introduce the following vacuum to the pseudo-scalar meson correlation function:
\begin{equation}
\Pi _{(\mu)}(q)=i\int d^{4}xe^{iq\cdot x}\langle \pi(q)|\mathcal{T}\{\eta_{\Lambda_b
}(x)\bar{\eta}_{\Sigma_b^{(*)}(\mu) }(0)\}|0\rangle,  \label{eq:CorrF2}
\end{equation}
where the on-shell $\pi$-meson state is represented by $\langle \pi(q)|$  with  momentum $q$, 
$\eta_{\Sigma_b^{(*)}(\mu)}$ is used to represent the interpolating current of $\Sigma_b^{\pm}(\Sigma_b^{*\pm})$, and $\eta_{\Lambda_b}$  is the interpolating current for the $\Lambda_b$ baryon having $J=\frac{1}{2}$. The interpolating fields  for the $J=\frac{1}{2} $-particles are given as
\begin{eqnarray}\label{Eq:Current1a}
\label{eapo03}
\eta_{\Lambda_b} &=& \frac{1}{\sqrt6}\epsilon^{abc}\Big\{
2\big(u_a^T C d_b \big) \gamma^5b_c +2 \beta \big(u_a^T C \gamma^5
d_b \big)b_c +\big(u_a^T C b_b \big) \gamma^5 d_c + \beta
\big(u_a^T C \gamma^5 b_b \big)d_c+
\big(b_a^T C d_b \big) \gamma^5 u_c \nonumber\\
&+& \beta
\big(b_a^T C \gamma^5 d_b \big)u_c\Big\},
\end{eqnarray}
and
\begin{eqnarray}\label{Eq:Current1a}
\label{eapo03}
\eta_{\Sigma_b} &=& \frac{1}{\sqrt2}\epsilon^{abc}\Big\{
\big(q_a^T C b_b \big) \gamma^5q_c + \beta \big(q_a^T C \gamma^5
b_b \big)q_c -\big(b_a^T C q_b \big) \gamma^5 q_c - \beta
\big(b_a^T C \gamma^5 q_b \big)q_c\Big\}.
\end{eqnarray}
For the states with $J=\frac{3}{2} $, we have:
\begin{eqnarray}
\label{Eq:Current1b}
\eta_{{\Sigma_b^{*}}{\mu}} = \sqrt{\frac{1}{3}} \epsilon^{abc} \Big\{ (q_a^{T} C \gamma_\mu q_b) b_c + (q_a^{T} C
\gamma_\mu b_b) q_c + (b_a^{T} C \gamma_\mu q_b) q_c \Big\}~.
\end{eqnarray}
In above equations, $q$ is the $u(d)$ quark field for $\Sigma_b^{(*)+}(\Sigma_b^{(*)-})$.
The indices $a$, $b$, and $c$ represent the colors, $C$ is the charge conjugation operator and $\beta$ is an arbitrary mixing parameter. This mixing parameter is introduced to include all the possible quark configurations in the interpolating currents considering the quantum numbers of the particles under considerations in order to write the  possible general forms of the interpolating currents for the particles with $J=\frac{1}{2}$. The case $\beta=-1$ corresponds to the Ioffe current. 

To obtain the sum rules for the strong coupling constants we start with the standard procedures of the QCD sum rules derivations. To obtain the physical or phenomenological  sides of  the desired sum rules, we insert complete sets of  the $\Sigma_b(\Sigma_b^{*})$ and $\Lambda_b$ baryons into the correlation function. As a result, we get 
\begin{eqnarray}
\Pi_{(\mu)} ^{\mathrm{Phys}}(p,q)&=&\frac{\langle 0|\eta_{\Lambda_b}|\Lambda_b (p,s)\rangle
}{p^{2}-m_{\Lambda_b}^{2}}\langle \pi(q)\Lambda_b(p,s)|\Sigma_b^{(*)}
(p^{\prime },s^{\prime })\rangle  \frac{\langle \Sigma_b^{(*)}(p^{\prime },s^{\prime
})|\bar{\eta}_{\Sigma_b^{(*)}{}{(\mu)}}|0\rangle }{p^{\prime
2}-m^{(*)2}}\nonumber\\
&+&\frac{\langle 0|\eta _{\Lambda_b}|\Lambda_b (p,s)\rangle
}{p^{2}-m_{\Lambda_b}^{2}}\langle \pi(q)\Lambda_b(p,s)|\Sigma^{(*)}_{b1}
(p^{\prime },s^{\prime })\rangle  \frac{\langle \Sigma^{(*)}_{b1}(p^{\prime },s^{\prime
})|\bar{\eta}_{\Sigma_b^{(*)}{}{(\mu)}}|0\rangle }{p^{\prime
2}-{m_{1}}^{(*)2}}
 +\ldots ,  \label{eq:SRDecay1a}
\end{eqnarray} 
\begin{eqnarray}
\Pi_{(\mu)} ^{\mathrm{Phys}}(p,q)&=&\frac{\langle 0|\eta_{\Lambda_b}|\Lambda_b(p,s)\rangle
}{p^{2}-m_{\Lambda_b}^{2}}\langle \pi(q)\Lambda_b(p,s)|\Sigma_b^{(*)}
(p^{\prime },s^{\prime })\rangle  \frac{\langle \Sigma_b^{(*)} (p^{\prime },s^{\prime
})|\bar{\eta}_{\Sigma_b^{(*)}{}{(\mu)}}|0\rangle }{p^{\prime
2}-m^{(*)2}}\nonumber\\
&+&\frac{\langle 0|\eta _{\Lambda_b}|\Lambda_b (p,s)\rangle
}{p^{2}-m_{\Lambda_b}^{2}}\langle \pi(q)\Lambda_b(p,s)|\Sigma^{(*)}_{b2}
(p^{\prime },s^{\prime })\rangle  \frac{\langle \Sigma^{(*)}_{b2}(p^{\prime },s^{\prime
})|\bar{\eta}_{\Sigma_b^{(*)}{}{(\mu)}}|0\rangle }{p^{\prime
2}-m_{2}^{(*)2}}
 +\ldots ,  \label{eq:SRDecay1b}
\end{eqnarray} 
where $p$ is the momentum of the $\Lambda_b$ baryon and  $p^{\prime}=p+q$ is the momentum of the considered $\Sigma_{b}^{(*)}$ and $\Sigma_{bi}^{(*)}$ initial states, with $i=1$ or $2$ indicating the $1P$ or $2S$ excited state. The dots at the ends of equations are used to represent the contributions of the higher states and the continuum.
It is well known that the physical (hadronic) side of the correlation function are complicated by the appearance of the contributions from the baryonic states of both positive and negative parities. Constructing the QCD sum rules for physical quantities 
free of the pollution from the unwanted (opposite) parity partners is of great importance (see Ref. \cite{Khodjamirian:2011jp} for more details). In our case, the hadronic side of the correlation function contains contributions from $1S$, $1P$ and $2S$  states at the same time. However, it is impossible to analytically solve the resultant coupled equations and separate different contributions from each other when three resonances are encountered. For this reason, in present work, we use the  ansatz that the hadronic side contains contributions either from $1S+1 P$ or $1S+2S$ states. By this way, we assume that  the observed states  $\Sigma_b(6097)^{\pm}$ to be either $1P$ or $2S$ excitations of   the corresponding ground-state baryons with $J=\frac{1}{2}$ or  $J=\frac{3}{2}$.  Then we separate the corresponding contributions of each state in each case. Naturally, such an assumption  brings some systematic uncertainties. However,  in order to estimate the order of systematic uncertainties due to this assumption it is also necessary to take simultaneously into account contributions of $1P$ and $2S$  states. In this case, we need to numerically solve the resultant three coupled equations. Analysis of this scenario lies beyond the scope of this work and we are planning to discuss this point in future, separately.

%However, as we said above if we consider the $1S+1 P+2 S$ scenario it is not possible to exactly solve the resultant coupled equations to estimate the order of %uncertainties coming from this assumption. Hence, estimation of the order of errors arising from this ansatz lies beyond the scope of this work. 

After using the matrix elements given in Eq.~(\ref{eq:Matrixelements1}) together with the following matrix elements, defined in terms of the decay constants, $ \lambda^{(*)} $, $ \lambda^{(*)}_{1} $, $ \lambda^{(*)}_{2} $ and $\lambda_{\Lambda_b} $,
\begin{eqnarray}
\langle 0|\eta_{\Sigma_b} |\Sigma_b(p',s)\rangle &=&\lambda u(p',s),
\nonumber \\
\langle 0|\eta_{\Sigma_b} |\Sigma_{b1}(p',s)\rangle
 &=&\lambda_{1}\gamma_5 u(p',s),
\nonumber \\
\langle 0|\eta_{\Sigma_b} |\Sigma_{b2}(p',s)\rangle
 &=&\lambda_{2} u(p',s),
 \nonumber \\
\langle 0|\eta_{\Lambda_b} |\Lambda_b(p,s)\rangle
 &=&\lambda_{\Lambda_b} u(p,s),
\label{eq:Res11}
\end{eqnarray}
%\
for the $J=\frac{1}{2}$-states and
\begin{eqnarray}
\langle 0|\eta_{\Sigma_b^{*}\mu } |\Sigma_b^{*}(p',s)\rangle &=&\lambda^{*}u_{\mu}(p',s),
\nonumber \\
\langle 0|\eta_{\Sigma_b^{*}\mu } |\Sigma^*_{b1}(p',s)\rangle
 &=&\lambda^{*}_1 \gamma_5u_{\mu}(p',s),
\nonumber \\
\langle 0|\eta_{\Sigma_b^{*}\mu } |\Sigma^{*}_{b2}(p',s)\rangle
 &=&\lambda^{*}_{2}u_{\mu}(p',s),
\label{eq:Res22}
\end{eqnarray}
%\
for the $J=\frac{3}{2}$-states, inside Eqs.~(\ref{eq:SRDecay1a}) and (\ref{eq:SRDecay1b}), and making the summations over spins using
\begin{eqnarray}\label{Dirac}
\sum_s  u (k,s)  \bar{u} (k,s) &= &(\!\not\!{k} + m),
\end{eqnarray}
\begin{eqnarray}\label{Rarita}
\sum_s  u_{\mu} (k,s)  \bar{u}_{\nu} (k,s) &= &-(\!\not\!{k} + m)\Big[g_{\mu\nu} -\frac{1}{3} \gamma_{\mu} \gamma_{\nu} - \frac{2k_{\mu}k_{\nu}}{3m^{2}} +\frac{k_{\mu}\gamma_{\nu}-k_{\nu}\gamma_{\mu}}{3m} \Big],
\end{eqnarray}
the results become
\begin{eqnarray}
\Pi ^{\mathrm{Phys}}(p,q)&=&\frac{%
g_{\Sigma_b \Lambda_b \pi}\lambda _{\Lambda_b}\lambda }{%
(p^{2}-m_{\Lambda_b}^{2})(p^{\prime }{}^{2}-m^{ 2})}(\slashed  q\slashed p\gamma_5+(m-m_{\Lambda_b})\slashed p \gamma_5) +\frac{g_{\Sigma_{b1}\Lambda_b\pi}\lambda_{\Lambda_b}%
\lambda_{1} }{(p^{2}-m_{\Lambda_b}^{2})(p^{\prime }{}^{2}-m_{1}^{2})}(\slashed q\slashed p\gamma_5-(m_1+m_{\Lambda_b})\slashed p\gamma_5) \nonumber\\
&+&\ldots , \label{eq:SRDecayPhys1a}
\end{eqnarray}%
\begin{eqnarray}
\Pi ^{\mathrm{Phys}}(p,q)&=&\frac{%
g_{\Sigma_b \Lambda_b\pi}\lambda _{\Lambda_b}\lambda }{%
(p^{2}-m_{\Lambda_b}^{2})(p^{\prime }{}^{2}-m^{ 2})}(\slashed q\slashed p\gamma_5+(m-m_{\Lambda_b})\slashed p \gamma_5) +\frac{%
g_{\Sigma_{b2} \Lambda_b \pi}\lambda_{\Lambda_b}\lambda_2}{%
(p^{2}-m_{\Lambda_b}^{2})(p^{\prime }{}^{2}-m_{2}^{ 2})}(\slashed q\slashed p\gamma_5+(m_2-m_{\Lambda_b})\slashed p\gamma_5)\nonumber\\
&+&\ldots ,\label{eq:SRDecayPhys1b}
\end{eqnarray}%
\begin{eqnarray}
\Pi _{\mu }^{\mathrm{Phys}}(p,q)&=&-\frac{g_{\Sigma_b^{*} \Lambda_b\pi}\lambda
_{\Lambda_b}\lambda}{(p^{2}-m_{\Lambda_b}^{2})(p^{\prime
2}-m^{*}{}^{2})}\Big[\frac{(m_{\Lambda_b}^2+2m_{\Lambda_b}m^{*}+m^{*}{}^2-m_{\pi}^2)}{6m^{*}}\slashed q \slashed p \gamma_{\mu}+\frac{(m_{\Lambda_b}^2-m_{\Lambda_b}m^{*}+m^{*}{}^2-m_{\pi}^2)m_{\Lambda_b}}{3m^{*}{}^2}\slashed q q_{\mu}\Big]\nonumber\\
& +&
\frac{g_{\Sigma_{b1}^{*} \Lambda_b\pi}\lambda
_{\Lambda_b}\lambda_{1}^{*}}{(p^{2}-m_{\Lambda_b}^{2})(p^{\prime }{}^{2}-%
m_{1}^{*}{}^{2})} \Big[\frac{(m_{\Lambda_b}^2-2m_{\Lambda_b}m_1^{*}+m_1^{*}{}^2-m_{\pi}^2)}{6 m_1^{*}}\slashed q \slashed p \gamma_{\mu}-\frac{(m_{\Lambda_b}^2+m_{\Lambda_b}m_1^{*}+m_1^{*}{}^2-m_{\pi}^2)m_{\Lambda_b}}{3 m_1^{*}{}^2}\slashed q q_{\mu}\Big]\notag \\
&+&\ldots,\label{eq:SRDecayPhys2a}
\end{eqnarray}
\begin{eqnarray}
\Pi _{\mu }^{\mathrm{Phys}}(p,q)&=&-\frac{g_{\Sigma_b^{*} \Lambda_b\pi}\lambda
_{\Lambda_b}\lambda}{(p^{2}-m_{\Lambda_b}^{2})(p^{\prime
2}-m^{*}{}^{2})}\Big[\frac{(m_{\Lambda_b}^2+2m_{\Lambda_b}m^{*}+m^{*}{}^2-m_{\pi}^2)}{6m^{*} }\slashed q \slashed p \gamma_{\mu}+\frac{(m_{\Lambda_b}^2-m_{\Lambda_b}m^{*}+m^{*}{}^2-m_{\pi}^2)m_{\Lambda_b}}{3m^{*}{}^2}\slashed q q_{\mu}\Big]\nonumber\\
& -&
\frac{g_{\Sigma_{b2}^{*} \Lambda_b\pi}\lambda
_{\Lambda_b}\lambda_{2}^{*}}{(p^{2}-m_{\Lambda_b}^{2})(p^{\prime }{}^{2}-%
m_{2}^{*}{}^{2})} \Big[\frac{(m_{\Lambda_b}^2+2m_{\Lambda_b}m_2^{*}+m_2^{*}{}^2-m_{\pi}^2)}{6 m_2^{*}}\slashed q \slashed p \gamma_{\mu}+\frac{(m_{\Lambda_b}^2-m_{\Lambda_b}m_2^{*}+m_2^{*}{}^2-m_{\pi}^2)m_{\Lambda_b}}{3m_2^{*}{}^2}\slashed q q_{\mu}\Big]\notag \\
&+&\ldots\label{eq:SRDecayPhys2b},
\end{eqnarray}%
where we only keep the terms that we use in the analyses and the dots in all the final results represent the contributions coming from  other structures as well as the higher states and  continuum. By applying the double Borel transformation with respect to $-p^2$ and $-p'^2$, we suppress the contributions of the higher states and the continuum, and after this process, the  Eqs.~(\ref{eq:SRDecayPhys1a})-(\ref{eq:SRDecayPhys2b}) become
\begin{eqnarray}
\tilde{\Pi} ^{\mathrm{Phys}}(p,q)&=&g_{\Sigma_b \Lambda_b\pi}\lambda _{\Lambda_b}\lambda
   e^{-m^{2}/M_{1}^{2}}e^{-m_{\Lambda_b}^{2}/M_{2}^{2}}[ %
\slashed q\slashed p\gamma _{5}+\left(
m-m_{\Lambda_b}\right) \slashed p\gamma _{5}]
+g_{\Sigma_{b1} \Lambda_b \pi}\lambda _{\Lambda_b}\lambda_{1}e^{-m_{1}^{2}/M_{1}^{2}}e^{-m_{\Lambda_b}^{2}/M_{2}^{2}}  \notag \\
&&\times [ \slashed q\slashed p\gamma _{5}-\left( m_{1}+m_{\Lambda_b}\right) \slashed p\gamma
_{5}]+\ldots  ,  \label{eq:CFunc1/2a}
\end{eqnarray}
\begin{eqnarray}
\tilde{\Pi} ^{\mathrm{Phys}}(p,q)&=&g_{\Sigma_b \Lambda_b \pi}\lambda _{\Lambda_b}\lambda
 e^{-m^{2}/M_{1}^{2}}e^{-m_{\Lambda_b}^{2}/M_{2}^{2}}[ %
\slashed q\slashed p\gamma _{5}+\left(
m-m_{\Lambda_b}\right) \slashed p\gamma _{5}]
+g_{\Sigma_{b2} \Lambda_b \pi}\lambda _{\Lambda_b}\lambda_2
   e^{-m_{2}^{2}/M_{1}^{2}}e^{-m_{\Lambda_b}^{2}/M_{2}^{2}} \notag \\
&&\times [ \slashed q\slashed p\gamma _{5}+\left(
m_{2}-m_{\Lambda_b}\right) \slashed p\gamma _{5}]+\ldots ,  \label{eq:CFunc1/2b}
\end{eqnarray}%
\begin{eqnarray}
\tilde{\Pi} _{\mu }^{\mathrm{Phys}}(p,q)&=&-g_{\Sigma_{b}^{*} \Lambda_b \pi}\lambda _{\Lambda_b}\lambda^{* }e^{-m^{*}{}%
^{2}/M_{1}^{2}}e^{-m_{\Lambda_b}^{2}/M_{2}^{2}}\Big[\frac{(m_{\Lambda_b}^2+2m_{\Lambda_b}m^{*}+m^{*}{}^2-m_{\pi}^2)}{6m^{*}}\slashed q \slashed p \gamma_{\mu}\nonumber\\
&+&\frac{(m_{\Lambda_b}^2-m_{\Lambda_b}m^{*}+m^{*}{}^2-m_{\pi}^2)m_{\Lambda_b}}{3m^{*}{}^2}\slashed q q_{\mu}\Big]
 +
g_{\Sigma_{b1}^{*} \Lambda_b\pi}\lambda _{\Lambda_b}\lambda_{1}^{*}  
 e^{-m_{1}^{*}{}^2/M_{1}^{2}}e^{-m_{\Lambda_b}^{2}/M_{2}^{2}}\nonumber\\
 &\times& \Big[\frac{(m_{\Lambda_b}^2-2m_{\Lambda_b}m_1^{*}+m_1^{*}{}^2-m_{\pi}^2)}{6 m_1^{*}}\slashed q \slashed p \gamma_{\mu}-\frac{(m_{\Lambda_b}^2+m_{\Lambda_b}m_1^{*}+m_1^{*}{}^2-m_{\pi}^2)m_{\Lambda_b}}{3 m_1^{*}{}^2}\slashed q q_{\mu}\Big]\notag \\
&+&\ldots,\label{eq:CFunc3/2a}
\end{eqnarray}
\begin{eqnarray}
\tilde{\Pi} _{\mu }^{\mathrm{Phys}}(p,q)&=&-g_{\Sigma_b^{*} \Lambda_b\pi}\lambda _{\Lambda_b}\lambda^{* }e^{-m^{*}{}%
^{2}/M_{1}^{2}}e^{-m_{\Lambda_b}^{2}/M_{2}^{2}} \Big[\frac{(m_{\Lambda_b}^2+2m_{\Lambda_b}m^{*}+m^{*}{}^2-m_{\pi}^2)}{6m^{*} }\slashed q \slashed p \gamma_{\mu}\nonumber\\
&+&\frac{(m_{\Lambda_b}^2-m_{\Lambda_b}m^{*}+m^{*}{}^2-m_{\pi}^2)m_{\Lambda_b}}{3m^{*}{}^2}\slashed q q_{\mu}\Big]-
g_{\Sigma_{b2}^{*} \Lambda_b\pi}\lambda _{\Lambda_b}\lambda^{*}_{2} e^{-m^{*}_{2}{}%
^{2}/M_{1}^{2}}e^{-m_{\Lambda_b}^{2}/M_{2}^{2}} \nonumber\\
& \times & \Big[\frac{(m_{\Lambda_b}^2+2m_{\Lambda_b}m_2^{*}+m_2^{*}{}^2-m_{\pi}^2)}{6 m_2^{*}}\slashed q \slashed p \gamma_{\mu}+\frac{(m_{\Lambda_b}^2-m_{\Lambda_b}m_2^{*}+m_2^{*}{}^2-m_{\pi}^2)m_{\Lambda_b}}{3m_2^{*}{}^2}\slashed q q_{\mu}\Big]\notag \\
&+&\ldots\label{eq:CFunc3/2b},
\end{eqnarray}%
where $M_1^2$ and $M_2^2$ are the corresponding Borel parameters to be fixed later. In the above equations, the  notation $\tilde{\Pi}_{(\mu) }^{\mathrm{Phys}}(p,q)$ is used to show the Borel transformed form of $\Pi_{(\mu)}^{\mathrm{Phys}}(p,q)$, and we use $q^2=m_{\pi}^2$. Among the presented Lorentz structures, to get the sum rules for the coupling constants, we choose the  $\slashed q \slashed p\gamma_5$ and $\slashed p \gamma_5$ for  $J=\frac{1}{2}$ scenarios. The  structures considered for the $J=\frac{3}{2}$ scenarios are the $\slashed q \slashed p\gamma_{\mu}$ and $\slashed q q_{\mu}$. For $J=\frac{3}{2}$ scenarios, the selected structures are free from the undesired spin-$\frac{1}{2}$ pollution.

Besides the physical sides of the calculations we need the theoretical or QCD sides of the desired sum rules obtained from the correlation function, Eq.~(\ref{eq:CorrF2}), via the operator product expansion (OPE). To this end, the explicit forms of the interpolating currents are placed in the correlator and possible contractions are made between the quark fields using Wick's theorem. As a results of these contractions, we obtain the outcomes in terms of the heavy- and light-quark propagators. There also appear terms containing the matrix elements of the quark-gluon field operators between vacuum and $\pi$-meson states having the common form  $\langle \pi(q)|\bar{q}(x)\Gamma G_{\mu\nu} q(y)|0\rangle$ or $\langle \pi(q)|\bar{q}(x)\Gamma q(y)|0\rangle$. Their explicit expressions are given in terms of the $\pi$-meson distribution amplitudes (DAs)~(see Refs.~\cite{Belyaev:1994zk,Ball:2004ye,Ball:2004hn}). The $\Gamma$ and $G_{\mu\nu}$ denote the full set of Dirac matrices and the gluon field strength tensor, respectively. Using these matrix elements, one gets the nonperturbative parts contributing to the results in coordinate space. We then carry out the calculations in the momentum space and apply a double Borel transformation over the same variables as the physicsl sides.  After applying the continuum subtraction procedure, the coefficients of same Lorentz structures as in the physical sides are considered, and the matching of these coefficients  from both sides leads to the QCD sum rules for the strong coupling constants under question. Representing the Borel transformed results of the QCD sides with $\tilde{\Pi}_{1}^{(*)\mathrm{OPE}}$ and $\tilde{\Pi}_{2}^{(*)\mathrm{OPE}}$, we can depict the mentioned matches as follows:   
\begin{eqnarray}
&&g_{{\Sigma_b}\Lambda_b\pi} \lambda_{\Lambda_b } \lambda e^{-\frac{m^2}{M_1^2}}e^{-\frac{m_{\Lambda_b}}{M_2^2}}+g_{\Sigma_{b1}\Lambda_b \pi}\lambda_{\Lambda_b } \lambda_{1} e^{-\frac{m_{1}^{2}}{M_{1}^{2}}}e^{-\frac{m_{\Lambda_b}}{M_2^2}}=\tilde{\Pi}_1^{\mathrm{OPE}}, \nonumber\\
&&g_{{\Sigma}\Lambda_b \pi} \lambda_{\Lambda_b } \lambda  e^{-\frac{m^2}{M_1^2}} e^{-\frac{m_{\Lambda_b}}{M_2^2}}(m-m_{\Lambda_b})-g_{\Sigma_{b1}\Lambda_b\pi}\lambda_{\Lambda_b } \lambda_{1} e^{-\frac{
m_{1}^{2}}{M_{1}^{2}}}e^{-\frac{m_{\Lambda_b}}{M_2^2}}(m_{1}+m_{\Lambda_b})=\tilde{\Pi}_2^{\mathrm{OPE}}\label{eq:couplingpair1/2a},
\end{eqnarray}
\begin{eqnarray}
&&g_{{\Sigma}_b\Lambda_b\pi} \lambda_{\Lambda_b } \lambda  e^{-\frac{m^2}{M_1^2}}e^{-\frac{m_{\Lambda_b}}{M_2^2}}+g_{\Sigma_{b}\Lambda_b\pi}\lambda_{\Lambda_b } \lambda_{1} e^{-\frac{m_{1}^{2}}{M_{1}^{2}}}e^{-\frac{m_{\Lambda_b}}{M_2^2}} =\tilde{\Pi}_1^{\mathrm{OPE}},\nonumber\\
&&g_{\Sigma_b \Lambda_b\pi}  \lambda_{\Lambda_b } \lambda e^{-\frac{m^2}{M_1^2}}e^{-\frac{m_{\Lambda_b}}{M_2^2}}(m-m_{\Lambda_b})+g_{\Sigma_{b2}\Lambda_b\pi}\lambda_{\Lambda_b } \lambda_{2} e^{-\frac{m_{2}^{2}}{M_{1}^{2}}}e^{-\frac{m_{\Lambda_b}}{M_2^2}}(m_2-m_{\Lambda_b})=\tilde{\Pi}_2^{\mathrm{OPE}}\label{eq:couplingpair1/2b},
\end{eqnarray}
\begin{eqnarray}
&-&g_{\Sigma_b^{*}\Lambda_b\pi}  \lambda_{\Lambda_b} \lambda^{*} \frac{[(m^{*}+m_{\Lambda_b})^2-m_{\pi}^2]}{6m^{*}}e^{-\frac{m^{*}{}^2}{M_1^2}}e^{-\frac{m_{\Lambda_b}}{M_2^2}}+g_{\Sigma_{b1}^{*}\Lambda_b\pi}\lambda_{\Lambda_b } \lambda_{1}^{*} \frac{[m_{1}^{*}-m_{\Lambda_b})^2-m_{\pi}^2]}{6m_{1}^{*}}e^{-\frac{m_{1}^{*}{}^{2}}{M_{1}^{2}}}e^{-\frac{m_{\Lambda_b}}{M_2^2}}=\tilde{\Pi}_{1}^{*}{}^{\mathrm{OPE}},\nonumber\\
&-&g_{\Sigma_b^{*}\Lambda_b\pi}  \lambda_{\Lambda_b } \lambda^{*} \frac{[m^{*}{}^2+m_{\Lambda_b}^2-m^{*} m_{\Lambda_b}-m_{\pi}^2]m_{\Lambda_b}}{3m^{*}{}^2}e^{-\frac{m^{*}{}^2}{M_1^2}}e^{-\frac{m_{\Lambda_b}}{M_2^2}}-g_{\Sigma_{b1}^{*}\Lambda_b\pi}\lambda_{\Lambda_b } \lambda_{1}^{*}
\frac{[m_{1}^{*}{}^2+m_{\Lambda_b}^2+m_{1}^{*}m_{\Lambda_b}-m_{\pi}^2]m_{\Lambda_b}}{3m_{1}^{*}{}^2} \nonumber\\
&\times & e^{-\frac{m_{1}^{*}{}^{2}}{M_{1}^{2}}}e^{-\frac{m_{\Lambda_b}}{M_2^2}}=\tilde{\Pi}_{2}^{*}{}^{\mathrm{OPE}}\label{eq:couplingpair3/2a},
\end{eqnarray}
\begin{eqnarray}
&-&g_{\Sigma_b^{*}\Lambda_b\pi}  \lambda_{\Lambda_b}\lambda^{*} \frac{[(m^{*}+m_{\Lambda_b})^2-m_{\pi}^2]}{6m^{*}} e^{-\frac{m^{*}{}^2}{M_1^2}}e^{-\frac{m_{\Lambda_b}}{M_2^2}}-g_{\Sigma_{b2}^{*}\Lambda_b\pi} \lambda_{\Lambda_b }\lambda^{*}_{2} \frac{[(m^{*}_{2}+m_{\Lambda_b})^2-m_{\pi}^2]}{6m^{*}_{2}} e^{-\frac{m^{*}_{2}{}^{2}}{M_{1}^{2}}}e^{-\frac{m_{\Lambda_b}}{M_2^2}}=\tilde{\Pi}_{1}^{*}{}^{\mathrm{OPE}},\nonumber\\
&-&g_{\Sigma_b^{*}\Lambda_b\pi}  \lambda_{\Lambda_b } \lambda^{*} \frac{[m^{*}{}^2+m_{\Lambda_b}^2-m^{*} m_{\Lambda_b}-m_{\pi}^2]m_{\Lambda_b}}{3m^{*}{}^2}e^{-\frac{m^{*}{}^2}{M_1^2}}e^{-\frac{m_{\Lambda_b}}{M_2^2}}-g_{\Sigma^{*}_{b2}\Lambda_b\pi}\lambda_{\Lambda_b } \lambda^{*}_{2}
\frac{[m^{*}_{2}{}^2+m_{\Lambda_b}^2-m^{*}_{2} m_{\Lambda_b}-m_{\pi}^2]m_{\Lambda_b}}{3m^{*}_{2}{}^2} \nonumber\\
&\times & e^{-\frac{m^{*}_{2}{}^{2}}{M_{1}^{2}}}e^{-\frac{m_{\Lambda_b}}{M_2^2}}=\tilde{\Pi}_{2}^{*}{}^{\mathrm{OPE}}\label{eq:couplingpair3/2b},
\end{eqnarray}
where $ \tilde{\Pi}_{1}^{\mathrm{OPE}}(\tilde{\Pi}_{1}^{*}{}^{\mathrm{OPE}})$ and $ \tilde{\Pi}_{2}^{\mathrm{OPE}}(\tilde{\Pi}_{2}^{*}{}^{\mathrm{OPE}})$ represent the Borel transformed coefficients of the $\slashed q \slashed p\gamma_5(\slashed q \slashed p\gamma_{\mu})$ and $\slashed p \gamma_5(\slashed q q_{\mu})$ structures for the $J=\frac{1}{2}(\frac{3}{2})$ cases. The procedures of the calculations of these functions and their expressions  are very lengthy. Hence, in appendix, we briefly show how we calculate these functions and give only the explicit form of the $\tilde{\Pi}_1^{*\mathrm{OPE}}$ function for  the $\Sigma_b(6097)^+\rightarrow\Lambda_b^0 \pi^+$ transition as an illustration.

The QCD sum rules for the coupling constants are obtained from the numerical solutions of the equation pairs given in the Eqs.~(\ref{eq:couplingpair1/2a}) and (\ref{eq:couplingpair1/2b}) for the $J=\frac{1}{2}$ scenarios and the Eqs.~(\ref{eq:couplingpair3/2a}) and (\ref{eq:couplingpair3/2b}) for the $J=\frac{3}{2}$ scenarios. 

The calculations for the coupling constants require some input parameters presented in Table~\ref{tab:Param}. %
\begin{table}[tbp]
%\rowcolors{1}{lightgray}{white}
\begin{tabular}{|c|c|}
\hline\hline
Parameters & Values \\ \hline\hline
$m_{\Sigma_b^{+}}$                       & $5811.3±1.9~\mathrm{MeV}$ \cite{Tanabashi2018}\\
$m_{\Sigma_b^{-}}$                       & $5815.5\pm 1.8~\mathrm{MeV}$ \cite{Tanabashi2018}\\
$m_{\Sigma_b^{*}{}^+}$                   & $5832.1\pm1.9~\mathrm{MeV}$ \cite{Tanabashi2018}\\
$m_{\Sigma_b^{*}{}^-}$                   & $5835.1\pm1.9~\mathrm{MeV}$ \cite{Tanabashi2018}\\
$m_{b}$                                  & $4.18^{+0.04}_{-0.03}~\mathrm{GeV}$ \cite{Tanabashi2018}\\
$m_{d}$                                  & $4.7^{+0.5}_{-0.3}~\mathrm{MeV}$ \cite{Tanabashi2018}\\
$ \lambda_{\Lambda_b}$                   & $(3.85\pm 0.56)\times 10^{-2}~\mathrm{GeV}^3$ \cite{Azizi:2008ui}\\
$\langle \bar{q}q \rangle (1\mbox{GeV})$ & $(-0.24\pm 0.01)^3$ $\mathrm{GeV}^3$ \cite{Belyaev:1982sa}  \\
$\langle \bar{s}s \rangle $              & $0.8\langle \bar{q}q \rangle$ \cite{Belyaev:1982sa} \\
$m_{0}^2 $                               & $(0.8\pm0.1)$ $\mathrm{GeV}^2$ \cite{Belyaev:1982sa}\\
$\langle g_s^2 G^2 \rangle $             & $4\pi^2 (0.012\pm0.004)$ $~\mathrm{GeV}
^4 $\cite{Belyaev:1982cd}\\
\hline\hline
\end{tabular}%
\caption{Some input parameters used in the calculations of the coupling constants and the masses.}
\label{tab:Param}
\end{table}
Since the masses of the considered baryons are close to each other, we choose 
\begin{eqnarray}
M_1^2=M_2^2=2M^2~~~~~\mbox{obtained from}~~~~~M^2=\frac{M_1^2 M_2^2}{M_1^2+M_2^2}.
\end{eqnarray}
As is seen from the equations, Eqs.~(\ref{eq:couplingpair1/2a})-(\ref{eq:couplingpair3/2b}), for the analyses of the considered coupling constants we also need the mass values of the considered baryons, and their decay constants. To obtain the masses and the decay constants we consider the following correlation function:
\begin{equation}
T_{(\mu \nu)}(q)=i\int d^{4}xe^{iq\cdot x}\langle 0|\mathcal{T}\{\eta_{\Sigma_b^{(*)}(\mu)}(x){\bar{\eta}_{\Sigma_b^{(*)}(\nu)}}(0)\}|0\rangle ,  \label{eq:CorrF1}
\end{equation}        
where the current $\eta_{\Sigma_b^{(*)}(\mu)}$ corresponds to the considered $J=\frac{1}{2}(\frac{3}{2})$ state, composed of the quark fields  regarding the related quantum numbers. The sub-index $\Sigma_b$ is used to represent one of the states, $\Sigma_b^{\pm}$ having spin $\frac{1}{2}$ and $\Sigma_b^{*\pm}$ having $J=\frac{3}{2}$. To determine the masses of the $\Sigma_b^{(*)}$ states, we again consider two assumptions for each of the above-mentioned baryons, and four different QCD sum rules  are obtained. For this purpose, the interpolating currents given in  Eqs.~(\ref{Eq:Current1a}) and (\ref{Eq:Current1b}) are used.

In the two-point QCD sum rule method for mass, one again follows two ways in the calculation of the corresponding correlator. The first one includes the calculation of the correlator in terms of the hadronic degrees of freedom and therefore it  is called as the physical or the phenomenological side. For this purpose, the interpolating fields are treated as the operators creating or annihilating the states under consideration. Insertion of  complete sets of hadronic states having the same quantum numbers of the hadrons under question results in

\begin{eqnarray}
T_{(\mu\nu)}^{\mathrm{Phys}}(q)&=&\frac{\langle 0|\eta_{\Sigma_b^{(*)}(\mu )} |\Sigma_b^{(*)}(q,s)\rangle \langle \Sigma_b^{(*)} (q,s)|\bar{\eta}_{\Sigma_b^{(*)}(\nu)}|0\rangle}{m^{(*)}{}^{2}-q^{2}}
+\frac{\langle 0|\eta_{\Sigma_b^{(*)}(\mu )} |\Sigma_{b1}^{(*)}(q,s)\rangle \langle \Sigma_{b1}^{(*)}(q,s)|\bar{\eta}_{\Sigma_b^{(*)}(\nu)}|0\rangle}{m_{1}^{(*)}{}^{2}-q^{2}}
+\ldots,
\label{eq:phys1a}
\end{eqnarray}
and
\begin{eqnarray}
T_{(\mu\nu)}^{\mathrm{Phys}}(q)&=&\frac{\langle 0|\eta_{\Sigma_b^{(*)}(\mu) } |\Sigma_b^{(*)}(q,s)\rangle \langle \Sigma_b^{(*)}(q,s)|\bar{\eta}_{\Sigma_b^{(*)}(\nu)}|0\rangle}{m^{(*)}{}^{2}-q^{2}}
+\frac{\langle 0|\eta_{\Sigma_b^{(*)}(\mu) } |\Sigma_{b2}^{(*)}(q,s)\rangle \langle \Sigma_{b2}^{(*)}(q,s)|\bar{\eta}_{\Sigma_b^{(*)}(\nu)}|0\rangle}{m_2^{(*)}{}^2-q^{2}}
+\ldots,
\label{eq:phys1b}
\end{eqnarray}
where  Eq.~(\ref{eq:phys1a}) is obtained for the $1P$ excitation and  Eq.(\ref{eq:phys1b}) is for the $2S$ excitation scenarios, respectively, with $m^{(*)}$, $m_1^{(*)}$, and $m_2^{(*)}$ being the masses of the $1S$, $1P$, and $2S$ excited states of each considered $\Sigma_b^{(*)}$ baryons whose one-particle states are represented by $\vert \Sigma_b^{(*)} \rangle$, $\vert \Sigma_{b1}^{(*)}\rangle$, and $\vert \Sigma_{b2}^{(*)}\rangle$, respectively. The dots represent contributions of the higher states and the continuum. As seen from the last equations, these calculations also require the matrix elements given in the Eqs.~(\ref{eq:Res11}) and (\ref{eq:Res22}). In these calculations again, the ground state and its  $1P$ and $2S$  excitations are notated by $\Sigma_b(\Sigma_b^{*})$, $\Sigma_{b1}(\Sigma_{b1}^{*})$, and $\Sigma_{b2}(\Sigma_{b2}^{*})$ for corresponding $J=\frac{1}{2}(\frac{3}{2})$ baryon, respectively, and $\lambda(\lambda^{*})$,  $\lambda_{1}(\lambda^{*}_{1})$ and $\lambda_{2}(\lambda^{*}_{2})$ are their corresponding decay constants. After the usage of expressions for the matrix elements and  using the summation relations for spinors $u(q,s)$ and $u_{\mu}(q,s)$ given in Eqs~(\ref{Dirac}) and (\ref{Rarita}), the physical sides for the $J=\frac{1}{2}$ cases are obtained as
\begin{eqnarray}\label{PhyssSide1a}
T^{\mathrm{Phys}}(q)&=&\frac{\lambda^{2}(\!\not\!{q}+m)}{m^{2}-q^2}+\frac{{\lambda_{1}}^{2}(\!\not\!{q}-m_{1})}{m_{1}^{2}-q^2}+\ldots,
\end{eqnarray}
and
\begin{eqnarray}\label{PhyssSide1b}
T^{\mathrm{Phys}}(q)&=&\frac{\lambda^{2}(\!\not\!{q}+m)}{m^{2}-q^2}+\frac{\lambda_{2}^{2}(\!\not\!{q}+m_{2})}{m_{2}^{2}-q^2}+\ldots.
\end{eqnarray}
The similar steps give the results for the $J=\frac{3}{2}$ cases as
\begin{eqnarray}\label{PhyssSide}
T_{\mu\nu}^{\mathrm{Phys}}(q)&=&-\frac{\lambda^{*}{}^{2}}{q^{2}-m^{*}{}^{2}}(\!\not\!{q} + m^{*})\Big[g_{\mu\nu} -\frac{1}{3} \gamma_{\mu} \gamma_{\nu} - \frac{2q_{\mu}q_{\nu}}{3m^{*}{}^{2}} +\frac{q_{\mu}\gamma_{\nu}-q_{\nu}\gamma_{\mu}}{3m^{*}} \Big]\nonumber \\
&-&\frac{\lambda_{1}^{*}{}^{2}}{q^{2}-m_{1}^{*}{}^{2}}(\!\not\!{q} - m_{1}^{*})\Big[g_{\mu\nu} -\frac{1}{3} \gamma_{\mu} \gamma_{\nu} - \frac{2q_{\mu}q_{\nu}}{3m_{1}^{*}{}^{2}} +\frac{q_{\mu}\gamma_{\nu}-q_{\nu}\gamma_{\mu}}{3m_{1}^{*}} \Big]+\ldots,
\end{eqnarray}
and
\begin{eqnarray}\label{PhyssSide}
T_{\mu\nu}^{\mathrm{Phys}}(q)&=&-\frac{\lambda^{*}{}^{2}}{q^{2}-m^{*}{}^{2}}(\!\not\!{q} + m^{*})\Big[g_{\mu\nu} -\frac{1}{3} \gamma_{\mu} \gamma_{\nu} - \frac{2q_{\mu}q_{\nu}}{3m^{*}{}^2} +\frac{q_{\mu}\gamma_{\nu}-q_{\nu}\gamma_{\mu}}{3m^{*}} \Big]\nonumber\\&-&\frac{\lambda_{2}^{2}}{q^{2}-m^{*}_{2}{}^{2}}(\!\not\!{q} + m^{*}_{2})\Big[g_{\mu\nu} -\frac{1}{3} \gamma_{\mu} \gamma_{\nu} - \frac{2q_{\mu}q_{\nu}}{3m^{*}_{2}{}^{2}} +\frac{q_{\mu}\gamma_{\nu}-q_{\nu}\gamma_{\mu}}{3m^{*}_{2}} \Big]+\ldots.
\end{eqnarray}

As already mentioned, we need to follow a second way to calculate the same correlation function, Eq.~(\ref{eq:CorrF1}), which proceeds in terms of the quark and gluon degrees of freedom. For this side of the calculation, we exploit the explicit expressions of the interpolating currents and OPE. After making the possible contractions between the quark fields, the results turn into expressions containing heavy- and light-quark propagators. To attain the final results, the expressions of these quark propagators are used and  Fourier transformation from coordinate space to momentum space is applied to obtain the final form of the QCD sides. The results of this side are very lengthy; therefore, we will not give them here explicitly.

The calculations of the physical and the QCD sides are followed by the application of a Borel transformation to both sides, which suppresses the contributions coming from the higher states and continuum. Finally, the QCD sum rules are attained by matching the coefficients of the same Lorentz structures from both sides. In the present work, the mentioned structures are $ \!\not\!{q}$ and $I$ for the $J=\frac{1}{2}$ cases and $ \!\not\!{q}g_{\mu\nu}$ and $g_{\mu\nu}$ for the $J=\frac{3}{2}$ cases. While choosing the structures for the $J=\frac{3}{2}$ states, among the  various possibilities, the structures $ \!\not\!{q}g_{\mu\nu}$ and $g_{\mu\nu}$ are considered since the others contain the undesired contributions from the $J=\frac{1}{2}$ states as well.
After the application of the continuum subtraction, the obtained equation pairs are solved numerically for each state under consideration. These equations are given as  
\begin{eqnarray}
\lambda^{2} e^{-\frac{m^{2}}{M^{2}}}+\lambda_{1}^2(\lambda_{2}^{ 2}) e^{-\frac{m_{1}^2(m_{2}^{2})}{M^{2}}}&=&{\tilde{T}}_{1}^{\mathrm{OPE}},
\nonumber \\
m \lambda^{2} e^{-\frac{m^{2}}{M^{2}}}\mp m_{1}(m_{2}) \lambda_{1}^{2}(\lambda_{2}^{2}) e^{-\frac{m_{1}^{2}(m_{2}^2)}{M^{2}}}&=&{\tilde{T}}_{2}^{\mathrm{OPE}}.
\label{Eq:sumrule2}
\end{eqnarray}
In the second term of the second equation, we use the $-$ and  $+$ signs to represent the results for $1P$ excitation, $\Sigma_{b1}$, and $2S$ excitation, $\Sigma_{b2}$, respectively. To represent the expressions obtained in the QCD side of the calculations, we use  $\tilde{T}_i^{\mathrm{OPE}}$ with $i=1,2$, which are the coefficients of the structures $ \!\not\!{q}$ and $I$ for $J=\frac{1}{2}$ cases. To obtain the results corresponding to the $J=\frac{3}{2}$ cases, it suffices to make the changes $\lambda_{1} \rightarrow \lambda_{1}^{*}$, $\lambda_{2} \rightarrow \lambda^{*}_2{}$, $m_{1} \rightarrow m_{1}^{*}$, $m_2 \rightarrow m^{*}_{2}$ and ${\tilde{T}}_{i}^{\mathrm{OPE}} \rightarrow \tilde{T}^{*\mathrm{OPE}}_{i}$ where $\tilde{T}^{*\mathrm{OPE}}_{i}$ is used to represent the coefficients obtained from $\!\not\!{q}g_{\mu\nu}$  and $g_{\mu\nu}$ in the QCD side.

\begin{figure}[h!]
\begin{center}
\includegraphics[totalheight=5cm,width=7cm]{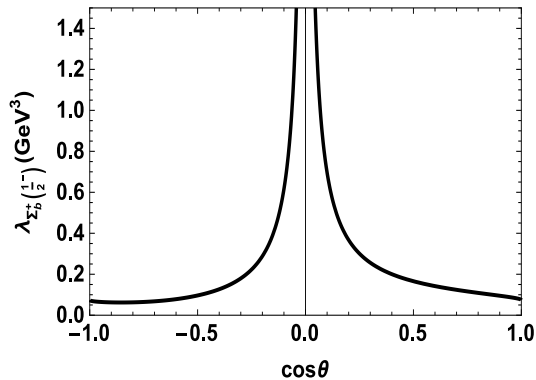}
\end{center}
\caption{ The dependence of the residue of $ \Sigma_b^+(\frac{1}{2}^-) $ state on $\cos\theta$ at average values of $M^2$ and  $s_0$. }
\label{fig1}
\end{figure}

\begin{figure}[h!]
\begin{center}
\includegraphics[totalheight=5cm,width=7cm]{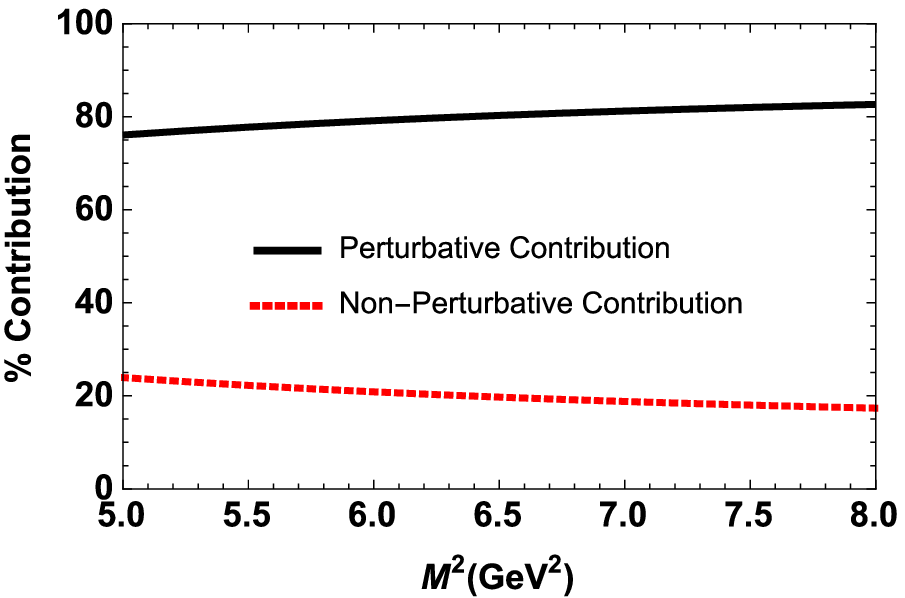}
\includegraphics[totalheight=5cm,width=7cm]{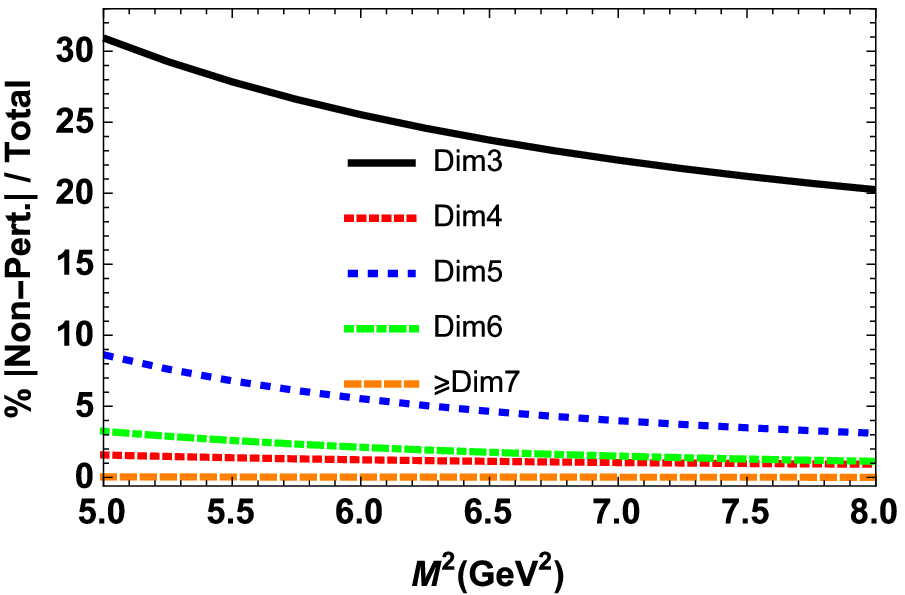}
\end{center}
\caption{ Various contributions to the OPE side of the mass sum rules for the $J=\frac{3}{2}$ case and the structure  $\!\not\!{q}g_{\mu\nu}$ on $M^2$ at average values of the $s_0$ and $\cos\theta$.}
\label{fig2}
\end{figure}

In the numerical analyses of the obtained results, we need some input parameters, which are presented in the Table~\ref{tab:Param}. The other ingredients of the sum rules are the three auxiliary parameters present in the results, namely the Borel parameter $M^2$, threshold parameter $s_0$, and an arbitrary parameter $\beta$. Note that the parameter $\beta$ belongs to the currents of the states with $J=\frac{1}{2}$. Their working regions are fixed via following some criteria of the QCD sum rule formalism. To decide on the relevant region for the Borel parameter, the convergence of the OPE calculation is considered. To satisfy this requirement, we demand a dominant perturbative contribution compared to the nonperturbative ones which helps us determine the lower limit of the Borel parameter. As for its upper limit, the criterion is the pole dominance. In technical language, for the upper band of the Borel window we require that 
\begin{eqnarray}
\frac{{\tilde{T}}_{i}^{(*)\mathrm{OPE}}(M^2,s0,\beta)}{{\tilde{T}}_{i}^{(*)\mathrm{OPE}}(M^2,\infty,\beta)}\geq \frac{1}{2},
\end{eqnarray}
while, for the lower band we demand that the perturbative part in each case exceeds the total nonperturbative contributions and the series of the corresponding OPE converge. 
From our analyses, we get this working interval as
\begin{eqnarray}
5~\mbox{GeV}^2\leq M^2\leq 8~\mbox{GeV}^2.
\end{eqnarray}
On the other hand, the threshold parameter, $s_0$,  is related to the energy of the first excited state of the considered state. Due to the lack of information about these excited states, this parameter is also determined considering  pole dominance condition as
\begin{eqnarray}
43~\mbox{GeV}^2\leq s_0 \leq 47~\mbox{GeV}^2.
\end{eqnarray}   
The parameter $\beta$ is determined from the analyses of the results searching for the region giving the least possible variation as a function of this parameter. This region is acquired via a parametric plot depicting the dependency of the result on $\cos\theta$, where $\beta=\tan\theta$. In figure \ref{fig1},  as an example, we plot the dependence of the residue of $ \Sigma_b^+(\frac{1}{2}^-) $ state on $\cos\theta$ at average values of $M^2$ and  $s_0$.  From this figure  and analyses of the obtained sum rules, the working region for the $\cos\theta$ is obtained as
\begin{eqnarray}
-1.0\leq\cos\theta\leq -0.3 ~~~~~\mbox{and} ~~~~~~0.3\leq \cos\theta\leq 1.0,
\end{eqnarray}
where the results demonstrate  small dependencies on the mixing parameter $\beta$. In order to see how the OPE sides of the mass sum rules converge, as an example, we show the dependence of the OPE side of the mass sum rule for the $J=\frac{3}{2}$ case and the structure  $\!\not\!{q}g_{\mu\nu}$ on $M^2$ at average values of the $s_0$ and $\cos\theta$ in figure \ref{fig2}. As is seen from this figure, the perturbative part constitutes the main contribution and the corresponding OPE series demonstrate a good convergence. 

With the usage of working intervals of the auxiliary parameters and the ones given in the Table~\ref{tab:Param}, the obtained masses and the decay constants are presented in Table~\ref{tab:results}. For the extraction of the masses for the considered excited states, the masses of corresponding ground-state baryons are used as inputs.   Note that  the central values  presented in this table are obtained at average values of $M^2$ and  $s_0$, i.e., $M^2=6.5~\mbox{GeV}^2$ and  $s_0=45~\mbox{GeV}^2$ as well as the average values of the  $\cos\theta$ in both the positive and negative sides.
\begin{table}[tbp]
%\rowcolors{1}{lightgray}{white}
\begin{tabular}{|c|c|c|}
\hline\hline
The state               & Mass (MeV) & Decay constant $\lambda~(\mbox{GeV}^3)$  \\
 \hline\hline
$ \Sigma_b^+(\frac{1}{2}^-)(1P)$     & $6091^{+197}_{-168} $ & $ 0.11^{+0.03}_{-0.03}$ \\
$ \Sigma_b^+(\frac{1}{2}^+)(2S)$     & $6091^{+197}_{-168} $ & $ 0.73^{+0.02}_{-0.04}$\\
$ \Sigma_b^-(\frac{1}{2}^-)(1P)$     & $6092^{+197}_{-168} $ & $ 0.11^{+0.06}_{-0.03} $\\
$ \Sigma_b^-(\frac{1}{2}^+)(2S)$     & $6092^{+197}_{-168} $ & $ 0.74^{+0.04}_{-0.02} $ \\
$ \Sigma_b^{*+}(\frac{3}{2}^-)(1P)$  & $6093^{+108}_{-123} $ & $ 0.068^{+0.010}_{-0.011}$ \\
$ \Sigma_b^{*+}(\frac{3}{2}^+)(2S)$  & $6093^{+108}_{-123} $ & $ 0.47^{+0.06}_{-0.02}$\\
$ \Sigma_b^{*-}(\frac{3}{2}^-)(1P)$  & $6095^{+107}_{-122} $ & $ 0.068^{+0.010}_{-0.011} $\\
$ \Sigma_b^{*-}(\frac{3}{2}^+)(2S)$  & $6095^{+107}_{-122} $ & $ 0.47^{+0.04}_{-0.02} $ \\
\hline\hline
\end{tabular}%
\caption{The results of the spectroscopic parameters obtained for the $1P$ and $2S$ excitations of the ground state $\Sigma_b^+$ and $\Sigma_b^-$ baryons with $J=\frac{1}{2}$ and $\Sigma_b^{*+}$ and $\Sigma_b^{*-}$ with $J^P=\frac{3}{2}$.}
\label{tab:results}
\end{table}
This table also contains the errors in the results coming from uncertainties existing in the input parameters and the uncertainties arising in the determination of the working windows for auxiliary parameters.

As seen from the table, although the mass results are consistent with that of the experimental observation given as $m_{\Sigma_b(6097)^-}=6098.0\pm1.7\pm0.5$~MeV and $m_{\Sigma_b(6097)^+}=6095.8\pm1.7\pm0.4$~MeV \cite{Aaij:2018tnn}, their central values are too close to indicate a deterministic information about the quantum numbers of the observed $\Sigma_b(6097)$ states. Therefore, for this purpose it would be much more helpful to resort to the results obtained for the decay widths. These decay widths are obtained from the usage of the results of strong coupling constant calculations with the application of the obtained mass and decay constant values. 

After getting the masses and decay constants, we turn our attention again to the strong coupling constant calculations in which the results of above spectroscopic parameters are used as inputs. In the strong coupling constant analyses we adopt the auxiliary parameters used in the calculations of masses and decay constants with one exception. The Borel parameter $M^2$ in these calculations is revisited, and, considering the OPE series convergence and the pole dominance conditions, its interval for the strong coupling constants is determined as 
\begin{eqnarray}
15~\mbox{GeV}^2 \leq M^2\leq 25~\mbox{GeV}^2.
\end{eqnarray}

The coupling constants attained from the QCD sum rule analyses are used to get the related decay widths for the $1P$ and the $2S$ excitations of the considered states. To this end, we use the decay width formulas for the $J=\frac{1}{2}$ cases given as:
\begin{eqnarray}
\Gamma \left( \Sigma_{b1}\rightarrow \Lambda_b\pi\right) =\frac{%
g_{\Sigma_{b1}\Lambda_b \pi}^{2}}{8\pi m_{1}^{2}}\left[ (m_{1}%
+m_{\Lambda_b})^{2}-m_{\pi}^{2}\right]  f(m_{1},m_{\Lambda_b},m_{\pi}),
\end{eqnarray}%
for $1P$ excitations and
\begin{eqnarray}
\Gamma \left( \Sigma_{b2}\rightarrow \Lambda_b\pi\right)  &=&%
\frac{g_{\Sigma_{b2}\Lambda_b\pi}^{2}}{8\pi m_{2}^{ 2}}\left[ (m_{2}-m_{\Lambda_b})^{2}-m_{\pi}^{2}\right]  f(m_2 ,m_{\Lambda_b},m_{\pi}),
\end{eqnarray}
for the $2S$ excitations, respectively. 
For the $J=\frac{3}{2}$ cases the respective decay-width equations are  
\begin{eqnarray}
\Gamma (\Sigma_{b1}^{*} &\rightarrow &\Lambda_b \pi)=\frac{%
g_{\Sigma_{b1}^{*}\Lambda_b \pi}^{2}}{24\pi m_{1}^{*}{}^{2}}\left[(m_{1}^{*}%
-m_{\Lambda_b})^{2}-m_{\pi}^{2}\right]  f^3(m_{1}^{*},m_{\Lambda_b},m_{\pi}),
\end{eqnarray}%
and
\begin{eqnarray}
\Gamma (\Sigma^{ * }_{b2} &\rightarrow &\Lambda_b\pi)=\frac{%
g_{\Sigma^{*}_{b2}\Lambda_b\pi}^{2}}{24\pi m^{*}_{2}{}^2}\left[ (m^{*}_{2}
+m_{\Lambda_b})^{2}-m_{\pi}^{2}\right]  f^3(m^{*}_{2},m_{\Lambda_b},m_{\pi}).
\end{eqnarray}
The function $f(x,y,z)$ present in the decay width equations is 
\begin{equation*}
f(x,y,z)=\frac{1}{2x}\sqrt{%
x^{4}+y^{4}+z^{4}-2x^{2}y^{2}-2x^{2}z^{2}-2y^{2}z^{2}}.
\end{equation*}

\begin{figure}[h!]
\begin{center}
\includegraphics[totalheight=5cm,width=7cm]{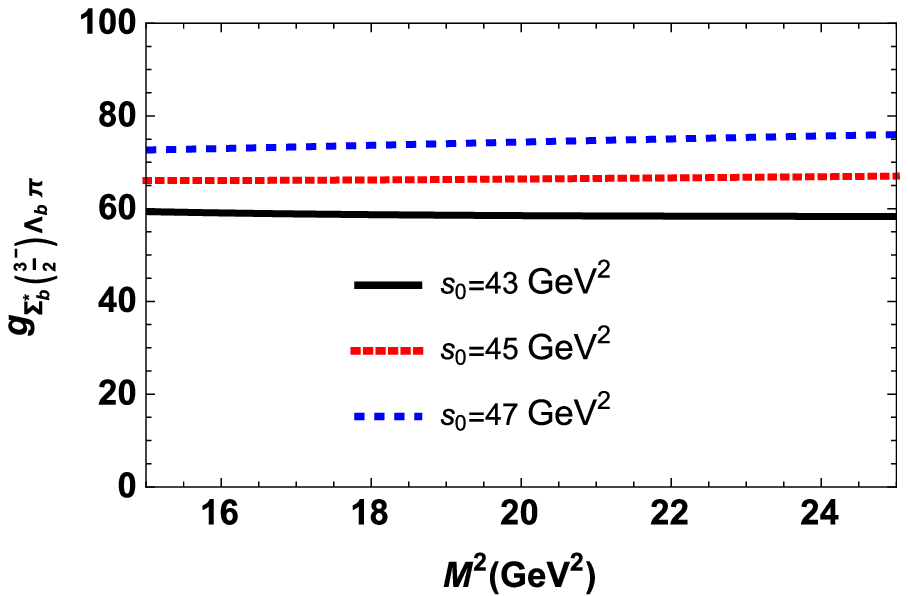}
\includegraphics[totalheight=5cm,width=7cm]{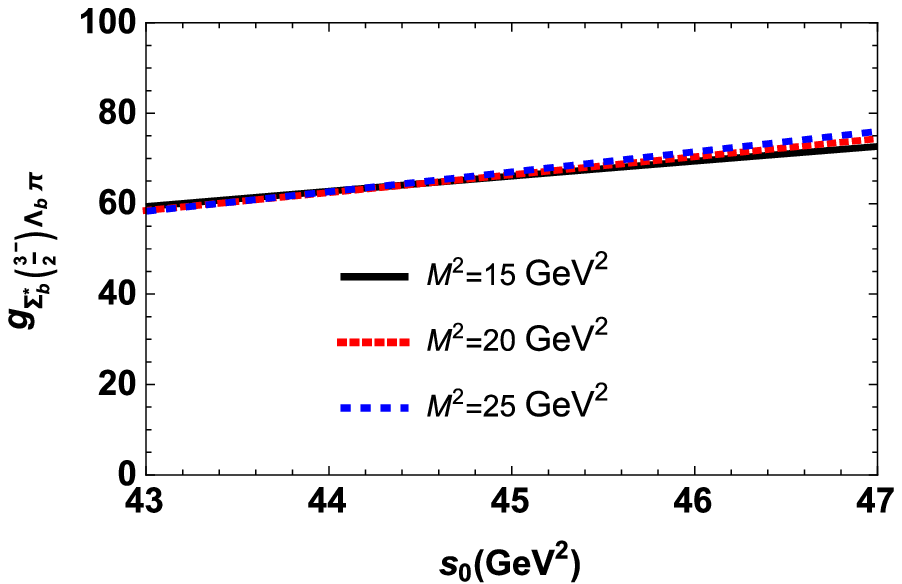}
\end{center}
\caption{ The dependence of the $ g_{\Sigma_b^{*+}(\frac{3}{2}^-)\Lambda_b\pi} $ on $ M^2 $ ($ s_0 $) at different fixed values of $ s_0 $ ($ M^2 $) and at average values of $\cos\theta$.}
\label{fig3}
\end{figure}

Table~\ref{tab:decayresults} presents the numerical results of the calculations for the coupling constants and decay widths. It can be seen from the table that our width results obtained for the scenario considering $\Sigma_b(6097)^{\pm}$ as the $1P$ excitations of the ground state $\Sigma_b^{*\pm}$ with $J^P=\frac{3}{2}^-$  are comparable to that of the experimental findings given as $\Gamma_{\Sigma_b(6097)^-}=28.9\pm4.2\pm 0.9$~MeV and $\Gamma_{\Sigma_b(6097)^+}=31.0\pm5.5\pm 0.7$~MeV~\cite{Aaij:2018tnn}. Note that the main uncertainties of the results for the couplings and masses belong to the variations of the results with respect to the variations of the continuum  threshold $ s_0 $ and the results show small dependencies on other auxiliary parameters as well as other input parameters. Figure \ref{fig3} shows the dependence of the $ g_{\Sigma_b^{*+}(\frac{3}{2}^-)\Lambda_b\pi} $ on $ M^2 $ ($ s_0 $) at different fixed values of $ s_0 $ ($ M^2 $) and at average values of $\cos\theta$. As is seen, the main source of uncertainties belongs to the variations of the continuum threshold $ s_0 $. 
\begin{table}[tbp]
%\rowcolors{1}{lightgray}{white}
\begin{tabular}{|c|c|c|}
\hline\hline
The state $B(J^P)$                &$g_{\Sigma_b\Lambda_b\pi}$&  $\Gamma~(\mbox{MeV})$  \\
 \hline\hline
$ \Sigma_b^+(\frac{1}{2}^-)(1P)$  &$1.4 \pm 0.3$    & $127.2 \pm 36.9 $ \\
$ \Sigma_b^+(\frac{1}{2}^+)(2S)$  &$9.1 \pm 2.0$    & $7.7 \pm 2.3 $\\
$ \Sigma_b^-(\frac{1}{2}^-)(1P)$  &$1.2 \pm 0.3$    & $85.4\pm 23.1 $\\
$ \Sigma_b^-(\frac{1}{2}^+)(2S)$  &$7.5 \pm 1.7$    & $5.4 \pm 1.6 $ \\

$ \Sigma_b^{*+}(\frac{3}{2}^-)(1P)$  &$67.7 \pm 14.9$  & $27.5 \pm 7.4 $ \\
$ \Sigma_b^{*+}(\frac{3}{2}^+)(2S)$  &$37.8 \pm 8.3$   & $5.7 \pm 1.6 $\\
$ \Sigma_b^{*-}(\frac{3}{2}^-)(1P)$  &$67.7 \pm 14.9$  & $28.1\pm 7.6 $\\
$ \Sigma_b^{*-}(\frac{3}{2}^+)(2S)$  &$37.8 \pm 8.3$   & $5.8 \pm 1.6 $ \\
\hline\hline
\end{tabular}%
\caption{The results of the coupling constants and the decay widths obtained for $1P$ and $2S$ excitations of the ground state $\Sigma_b^+$ and $\Sigma_b{}^-$  baryons having spin-$\frac{1}{2}$ and  $\Sigma_b^{*}{}^+$ and $\Sigma^{*}_b{}^-$ having spin-$\frac{3}{2}$.}
\label{tab:decayresults}
\end{table}

 At the end of this section we would like to compare our results for the masses and widths with the predictions of other approaches.  In Ref.\cite{Chen:2018vuc}, using the quasi-two-body method, the results for the masses were obtained as $6094$~MeV and $6098$~MeV for $\Sigma(3/2^-)$ and $\Sigma(5/2^-)$ sates, respectively, indicating the possibility for the particle $\Sigma_b(6097)$ having either $J^P=3/2^-$ or $5/2^-$. The result of the mass for $\Sigma(3/2^-)$ state is consistent with our predictions.  In the same reference the decay widths also considered and for the channel with final states $\Lambda_b \pi$ the results were presented as $35.2$~MeV and $35.8$~MeV for $\Sigma(3/2^-)$ and $\Sigma(5/2^-)$ sates, respectively, supporting their conclusion obtained from the mass calculations. The decay width calculations to the same final state for the strong decay of the P-wave $\Sigma_b$ baryon was also considered in Ref.~\cite{Wang:2018fjm} using chiral quark model which leaded to the results $32.3$~MeV and $31.4$~MeV for $J^P=3/2^-$ and $5/2^-$ considerations, respectively. Another study supporting the $\Sigma_b(6097)$ having either a $J^P=3/2^-$ or $5/2^-$ presented the decay widths as $14.56(14.19)$~MeV for the $\Sigma_{b}(6097)^-(\Sigma_{b}(6097)^+)$ for both $J^P=3/2^-$ and $5/2^-$ cases~\cite{Yang:2018lzg}. As is seen the results of \cite{Yang:2018lzg} for decay widths differ from our predictions and the experimental data, considerably. However, the predictions of \cite{Chen:2018vuc,Wang:2018fjm} are close to our predictions as well as the experimental results. The advantage of our predictions for the widths using the LCSR is that by combination of these predictions with the mass results we can exactly  assign the particles $\Sigma_b(6097)^{\pm}$ to be the $1P$ excitations of the ground state $\Sigma_b^{*\pm}$ baryons with quantum numbers $J^P=\frac{3}{2}^-$.

\section{Conclusion}
To investigate the properties of the recently observed $\Sigma_b(6097)^{\pm}$, the light cone QCD sum rule calculations were performed and the strong coupling constants for their transitions to $\Lambda_b^{0}\pi^{\pm}$ states were obtained. For the analyses, two possible cases, $J=\frac{1}{2}$ and $J=\frac{3}{2}$, were considered  and for each of them the  $1P$ and $2S$ excitations were taken into account. For each case, the considered decays were studied, and from the obtained strong coupling constants, the related decay widths were calculated. For the calculations of the strong coupling constants, the mass and the decay constant of each considered state with possible quantum numbers were required. To supply these quantities  we employed the two-point QCD sum rules. From the results of mass sum rule analyses, we obtained the  mass values as
$ m_{\Sigma_b^+(\frac{1}{2}:1P(2S))}=6091^{+197}_{-168} $~MeV, 
$ m_{\Sigma_b^-(\frac{1}{2}:1P(2S))}=6092^{+197}_{-168} $~MeV, 
$m_{\Sigma_b^{*+}(\frac{3}{2}:1P(2S))}=6093^{+108}_{-123}$~MeV, 
$m_{\Sigma_b^{*-}(\frac{3}{2}:1P(2S))}=6095^{+107}_{-122}$~MeV. 
As is seen, the central values obtained for masses are in consistency with the experimentally observed masses,  $m(\Sigma_b(6097)^-)=6098.0\pm1.7\pm0.5$~MeV and $m(\Sigma_b(6097)^+)=6095.8\pm1.7\pm0.4$~MeV~\cite{Aaij:2018tnn}. However it can be seen from these results, just looking at these mass values it is not possible to draw a conclusion about the quantum numbers of the states $\Sigma_b(6097)^{\pm}$. Because, the central values of the obtained results are close not only to the experimental results but also to each other, and this does not allow us to make a conclusive statement about the quantum numbers. Therefore, using them as input quantities in the calculations of the strong coupling constants, we attained the numerical values of the corresponding coupling constants and subsequently the related decay widths, which were the main focus of the present work. Our results for the decay widths obtained for the $J^P=\frac{3}{2}^-$ possibilities are $ \Gamma_{\Sigma_b^+(\frac{3}{2}^-)}=27.5 \pm 7.4 $~MeV and $ \Gamma_{\Sigma_b^-(\frac{3}{2}^-)}=28.1\pm 7.6 $~MeV, which are in accord with the observed widths of these states, i.e. $\Gamma(\Sigma_b(6097)^-)=28.9\pm 4.2\pm 0.9$~MeV and $\Gamma(\Sigma_b(6097)^+)=31.0\pm 5.5\pm 0.7$~MeV~\cite{Aaij:2018tnn}. These results support the states being $1P$ excitations of the ground state $\Sigma_b^{*\pm}$ with $J=\frac{3}{2}$.

%%%%%%%%%%%%%%%%%%%%%%%%%%%%%%%%%%%%%%%%%%%%%%%%%%%%%%%%%%%%

\section{Appendix: Some details of the calculations of the spectral densities  for the coupling constants}

Here we present some details of the calculations of the spectral densities used in the analyses of the strong coupling constants. After contracting out the quark fields in the QCD side,  there appear an expression in terms of the heavy and light quarks propagators as well as the matrix elements of the quark-gluon field operators between vacuum and  pseudoscalar meson states having the  forms  $\langle PS(q)|\bar{q}(x)\Gamma G_{\mu\nu} q(y)|0\rangle$ and $\langle PS(q)|\bar{q}(x)\Gamma q(y)|0\rangle$. These matrix elements are given in terms of the  pseudoscalar meson DAs~(see Refs.~\cite{Belyaev:1994zk,Ball:2004ye,Ball:2004hn}). For some details on the calculations of the spectral densities in QCD, we also refer the reader to  Ref. \cite{Nesterenko:1982gc}. 

 For the light and heavy quarks propagators we use
\begin{eqnarray}\label{prolight} 
S_q(x) \!\!\! &=& \!\!\! {i \rlap/x\over 2\pi^2 x^4} - {m_q\over 4 \pi^2 x^2} -
{\langle \bar q q \rangle \over 12} \left(1 - i {m_q\over 4} \rlap/x \right) -
{x^2\over 192} m_0^2 \langle \bar q q \rangle  \left( 1 -
i {m_q\over 6}\rlap/x \right) \nonumber \\
&&  - i g_s \int_0^1 du \left[{\rlap/x\over 16 \pi^2 x^2} G_{\mu \nu} (ux)
\sigma_{\mu \nu} - {i\over 4 \pi^2 x^2} u x^\mu G_{\mu \nu} (ux) \gamma^\nu
\right. \nonumber \\
&& \left.
 - i {m_q\over 32 \pi^2} G_{\mu \nu}(ux) \sigma^{\mu
 \nu} \left( \ln \left( {-x^2 \Lambda^2\over 4} \right) +
 2 \gamma_E \right) \right]~,
\end{eqnarray}
and
\begin{eqnarray}\label{proheavy} 
S_Q(x) \!\!\! &=& \!\!\! {m_Q^2 \over 4 \pi^2} {K_1(m_Q\sqrt{-x^2}) \over \sqrt{-x^2}} -
i {m_Q^2 \rlap/{x} \over 4 \pi^2 x^2} K_2(m_Q\sqrt{-x^2})\nonumber \\
&&-
ig_s \int {d^4k \over (2\pi)^4} e^{-ikx} \int_0^1
du \Bigg[ {\rlap/k+m_Q \over 2 (m_Q^2-k^2)^2} G^{\mu\nu} (ux)
\sigma_{\mu\nu}\nonumber \\
&& +
{u \over m_Q^2-k^2} x_\mu G^{\mu\nu} (ux)\gamma_\nu \Bigg]~,
\end{eqnarray}
where $\gamma_E $ is the Euler constant, $ G_{\mu \nu} $ is the gluon field strength tensor,  $\Lambda$ is the
scale parameter and $K_{\nu}$ in the heavy propagator denote the Bessel functions of the second kind.

After insertion of the light and heavy quarks propagators as well as the DAs of the pseudoscalar mesons we get the following generic term, as an example for the leading twist (see also \cite{Aliev:2011ufa}):                                                                                                                                                                   
\baeeq
\label{nolabel}
T=\int d^4x ~e^{ipx} \int_0^1 du~ e^{iuqx} f(u) \frac{K_\nu(m_Q\sqrt{-x^2})}{(\sqrt{-x^2})^n},
\eaeeq
where $f(u)$ denotes the leading DAs. We need to  perform the Fourier and Borel transformations as well as continuum subtraction on this expression. To this end, we use the integral representation of the modified Bessel function as
\baeeq
 K_\nu(m_Q\sqrt{-x^2})=\frac{\Gamma(\nu+1/2)2^\nu}{\sqrt{\pi}m_Q^\nu}\int_0^\infty dt~cos(m_Qt)\frac{(\sqrt{-x^2})^\nu}{(t^2-x^2)^{\nu+1/2}},
\eaeeq
which leads to
\baeeq
T=\int d^4x\int_0^1 du~e^{iPx}f(u)\frac{\Gamma(\nu+1/2)2^\nu}{\sqrt{\pi}m_Q^\nu}\int_0^\infty dt~cos(m_Qt)\frac{1}{(\sqrt{-x^2})^{n-\nu}(t^2-x^2)^{\nu+1/2}},
\eaeeq
where $P=p+uq$.  By transferring the calculations  into the Euclidean space and using the identity
\baeeq
\frac{1}{Z^n}=\frac{1}{\Gamma(n)}\int_0^\infty d\alpha~\alpha^{n-1} e^{-\alpha Z},
\eaeeq
we get
\baeeq
T=\frac{-i2^\nu}{\sqrt{\pi}m_Q^\nu\Gamma(\frac{n-\nu}{2})}\int_0^1 du f(u)\int_0^\infty dt~e^{im_Qt}\int_0^\infty dy ~y^{\frac{n-\nu}{2}-1}
\int_0^\infty dv ~v^{\nu-\frac{1}{2}}e^{-vt^2}\int d^4\tilde{x}e^{-i\tilde{P}\tilde{x}-y\tilde{x}^2-v\tilde{x}^2},\nnb\\
\eaeeq
where the sign $\sim$ refers to the vectors in Euclidean space. After performing the resultant Gaussian integral
 over four-$\tilde{x}$, we end up with
\baeeq
T=\frac{-i2^\nu\pi^2}{\sqrt{\pi}m_Q^\nu\Gamma(\frac{n-\nu}{2})}\int_0^1 du f(u)\int_0^\infty dt~e^{im_Qt}\int_0^\infty dy ~y^{\frac{n-\nu}{2}-1}
\int_0^\infty dv ~v^{\nu-\frac{1}{2}}e^{-vt^2}\frac{e^{-\frac{\tilde{P}^2}{4(y+v)}}}{(y+v)^2}.
\eaeeq
The next step is to perform the integration over $t$, which leads to
\baeeq
T=\frac{-i2^\nu\pi^2}{m_Q^\nu\Gamma(\frac{n-\nu}{2})}\int_0^1 du f(u) \int_0^\infty dy ~y^{\frac{n-\nu}{2}-1}
\int_0^\infty dv ~v^{\nu-1}e^{-\frac{m_Q^2}{4v}}\frac{e^{-\frac{\tilde{P}^2}{4(y+v)}}}{(y+v)^2}.
\eaeeq
Let us define the following new variables:
\baeeq
~~~~~~~~~~~~~~~~~~~~~~~~~\lambda=v+y,~~~~~~~~~~~~~\tau=\frac{y}{v+y}.
\eaeeq
Applying this, we obtain
\baeeq
T=\frac{-i2^\nu\pi^2}{m_Q^\nu\Gamma(\frac{n-\nu}{2})}\int_0^1 du f(u) \int d\lambda \int d\tau
 ~ \lambda^{\frac{n+\nu}{2}-3} \tau^{\frac{n-\nu}{2}-1}(1-\tau)^{\nu-1}e^{-\frac{m_Q^2}{4\lambda(1-\tau)}} e^{-\frac{\tilde{P}^2}{4\lambda}}
\eaeeq
Now, we perform the Double Borel transformation with respect to the $\tilde{p}^2$ and  $(\tilde{p}+\tilde{p})^2$ by the help of 
\baeeq
{\cal B}(M^2)e^{-\alpha p^2}=\delta(1/M^2-\alpha),
\eaeeq
which leads to
\baeeq
{\cal B}(M_1^2){\cal B}(M_2^2)T&=&\frac{-i2^\nu\pi^2}{m_Q^\nu\Gamma(\frac{n-\nu}{2})}\int_0^1 du f(u) \int d\lambda \int d\tau
 ~\lambda^{\frac{n+\nu}{2}-3} \tau^{\frac{n-\nu}{2}-1}(1-\tau)^{\nu-1}e^{-\frac{m_Q^2}{4\lambda(1-\tau)}}e^{-\frac{u(u-1)\tilde{q}^2}{4\lambda}}\nnb\\
&\times&
\delta(\frac{1}{M_1^2}-\frac{u}{4 \lambda})\delta(\frac{1}{M_2^2}-\frac{1-u}{4 \lambda}).
\eaeeq
In this step the integrals over $u$ and $\lambda$ are performed. As a result, we get
\baeeq
{\cal B}(M_1^2){\cal B}(M_2^2)T&=&\frac{-i2^\nu4^2\pi^2}{m_Q^\nu\Gamma(\frac{n-\nu}{2})} \int d\tau f(u_0) 
 \Bigg(\frac{M^2}{4}\Bigg)^{\frac{n+\nu}{2}}\tau^{\frac{n-\nu}{2}-1}(1-\tau)^{\nu-1}
e^{-\frac{m_Q^2}{M^2(1-\tau)}}e^{\frac{\tilde{q}^2}{M_1^2+M_2^2}},\nnb\\
\eaeeq
where, $u_0=\frac{M_2^2}{M_1^2+M_2^2}$ and $M^2=\frac{M_1^2M_2^2}{M_1^2+M_2^2}$. By the replacement $\tau=x^2$, we obtain
\baeeq
{\cal B}(M_1^2){\cal B}(M_2^2)T&=&\frac{-i2^{\nu+1}4^2\pi^2}{m_Q^\nu\Gamma(\frac{n-\nu}{2})} \int_0^1 dx f(u_0) 
 \Bigg(\frac{M^2}{4}\Bigg)^{\frac{n+\nu}{2}}x^{n-\nu-1}(1-x^2)^{\nu-1}
e^{-\frac{m_Q^2}{M^2(1-x^2)}}e^{\frac{\tilde{q}^2}{M_1^2+M_2^2}}.\nnb\\
\eaeeq
The last step is the  changing of the variable $\eta=\frac{1}{1-x^2}$ and using $q^2=m^2_{{\cal P}}$, which leads to  
\baeeq
{\cal B}(M_1^2){\cal B}(M_2^2)T&=&\frac{-i2^{\nu+1}4^2\pi^2}{m_Q^\nu\Gamma(\frac{n-\nu}{2})}  f(u_0) 
 \Bigg(\frac{M^2}{4}\Bigg)^{\frac{n+\nu}{2}}e^{-\frac{m^2_{{\cal P}}}{M_1^2+M_2^2}}\Psi\Bigg(\alpha,\beta,\frac{m_Q^2}{M^2}\Bigg),
\eaeeq
with
\baeeq
\Psi\Bigg(\alpha,\beta,\frac{m_Q^2}{M^2}\Bigg)=\frac{1}{\Gamma(\alpha)}\int_1^\infty d\eta e^{-\eta\frac{m_Q^2}{M^2}} \eta^{\beta-\alpha-1}(\eta-1)^{\alpha-1},
\eaeeq
where $\alpha=\frac{n-\nu}{2}$ and $\beta=1-\nu$.

In this stage, we discuss how  the contributions of the higher states and continuum  are subtracted. We consider the generic 
 form 
\baeeq
A=(M^2)^n f(u_0) \Psi\Bigg(\alpha,\beta,\frac{m_Q^2}{M^2}\Bigg).
\eaeeq
 We are going to find the spectral density corresponding to this generic term. As a first step,  we expand $f(u_0)$ as
\baeeq
f(u_0)=\Sigma a_ku_0^k,
\eaeeq
which leads to
\baeeq
A=\Bigg(\frac{M_1^2 M_2^2}{M_1^2+M_2^2}\Bigg)^n\Sigma a_k\Bigg(\frac{ M_2^2}{M_1^2+M_2^2}\Bigg)^k\frac{1}{\Gamma(\alpha)}\int_1^\infty d\eta 
e^{-\eta\frac{m_Q^2}{M^2}} \eta^{\beta-\alpha-1}(\eta-1)^{\alpha-1}.
\eaeeq
Now, we introduce the new variables, $\sigma_1=\frac{1}{M_1^2}$ and $\sigma_2=\frac{1}{M_2^2}$. As a result  we get
\baeeq
A&=&\Sigma a_k \frac{\sigma_1^k}{(\sigma_1+\sigma_2)^{n+k}}\frac{1}{\Gamma(\alpha)}\int_1^\infty d\eta 
e^{-\eta m_Q^2(\sigma_1+\sigma_2)} \eta^{\beta-\alpha-1}(\eta-1)^{\alpha-1}\nnb\\
&=&\Sigma a_k \frac{\sigma_1^k}{\Gamma(n+k)\Gamma(\alpha)}\int_1^\infty d\eta 
 e^{-\eta m_Q^2(\sigma_1+\sigma_2)}\eta^{\beta-\alpha-1}(\eta-1)^{\alpha-1}\int_0^\infty d\xi e^{-\xi(\sigma_1+\sigma_2)}\xi^{n+k-1}\nnb\\
&=&\Sigma a_k \frac{\sigma_1^k}{\Gamma(n+k)\Gamma(\alpha)}\int_1^\infty d\eta 
 \eta^{\beta-\alpha-1}(\eta-1)^{\alpha-1}\int_0^\infty d\xi \xi^{n+k-1}e^{-(\xi+\eta m_Q^2)(\sigma_1+\sigma_2)}\nnb\\
&=&\Sigma a_k \frac{(-1)^k}{\Gamma(n+k)\Gamma(\alpha)}\int_1^\infty d\eta 
 \eta^{\beta-\alpha-1}(\eta-1)^{\alpha-1}\int_0^\infty d\xi \xi^{n+k-1}\Bigg((\frac{d}{d\xi})^ke^{-(\xi+\eta m_Q^2)\sigma_1}\Bigg)e^{-(\xi+\eta m_Q^2)\sigma_2}.\nnb\\
\eaeeq
By applying the double Borel transformation with respect to $\sigma_1\rar\frac{1}{s_1}$ and $\sigma_2\rar\frac{1}{s_2}$, we obtain the following double spectral density
\baeeq
\rho(s_1,s_2)&=&\Sigma a_k \frac{(-1)^k}{\Gamma(n+k)\Gamma(\alpha)}\int_1^\infty d\eta 
 \eta^{\beta-\alpha-1}(\eta-1)^{\alpha-1}\int_0^\infty d\xi \xi^{n+k-1}\Bigg((\frac{d}{d\xi})^k\delta(s_1-(\xi+\eta m_Q^2))\Bigg)\nnb\\
&\times&\delta(s_2-(\xi+\eta m_Q^2)).
\eaeeq
By performing the integral over $\xi$,  we acquire the following expression for the double spectral density:
\baeeq
\rho(s_1,s_2)&=&\Sigma a_k \frac{(-1)^k}{\Gamma(n+k)\Gamma(\alpha)}\int_1^\infty d\eta 
 \eta^{\beta-\alpha-1}(\eta-1)^{\alpha-1}(s_1-\eta m_Q^2)^{n+k-1}\Bigg((\frac{d}{ds_1})^k\delta(s_2-s_1)\Bigg)\nnb\\
&\times&\theta(s_1-\eta m_Q^2),
\eaeeq
which can be written as
\baeeq
\rho(s_1,s_2)&=&\Sigma a_k \frac{(-1)^k}{\Gamma(n+k)\Gamma(\alpha)}\int_1^{s_1/m_Q^2} d\eta 
 \eta^{\beta-\alpha-1}(\eta-1)^{\alpha-1}(s_1-\eta m_Q^2)^{n+k-1}\Bigg((\frac{d}{ds_1})^k\delta(s_2-s_1)\Bigg).\nnb\\
\eaeeq
With the use of this double spectral density, the continuum subtracted correlation function in the Borel scheme corresponding to the generic term under consideration is written as
\baeeq
\Pi^{sub}=\int_{m_Q^2}^{s_0}ds_1\int_{m_Q^2}^{s_0}ds_2~\rho(s_1,s_2)e^{-s_1/M_1^2}e^{-s_2/M_2^2}.
\eaeeq
Now, we define the  new variables, $s_1=2 s v$ and $s_2=2 s(1- v)$. As a result, we obtain
\eAPP
\baeeq
\Pi^{sub}=\int_{m_Q^2}^{s_0}ds\int dv~\rho(s_1,s_2)(4s)e^{-2sv/M_1^2}e^{-2s(1-v)/M_2^2}.
\eaeeq
Inserting  the expression of  the above spectral density, one can immediately get
\baeeq
\Pi^{sub}&=&\Sigma a_k \frac{(-1)^k}{\Gamma(n+k)\Gamma(\alpha)}\int_{m_Q^2}^{s_0}ds\int dv\frac{1}{2^ks^k}\Bigg((\frac{d}{dv})^k\delta(v-1/2)\Bigg)\nnb\\
&\times&\int_1^{2sv/m_Q^2} d\eta ~
 \eta^{\beta-\alpha-1}(\eta-1)^{\alpha-1}(2sv-\eta m_Q^2)^{n+k-1}e^{-2sv/M_1^2}e^{-2s(1-v)/M_2^2}.
\eaeeq
Now, we perform the integration over $v$, which leads to the final form:
\baeeq
\Pi^{sub}&=&\Sigma a_k \frac{(-1)^k(-1)^k}{\Gamma(n+k)\Gamma(\alpha)}\int_{m_Q^2}^{s_0}ds\frac{1}{2^ks^k}\nnb\\
&\times&\Bigg[(\frac{d}{dv})^k\int_1^{2sv/m_Q^2} d\eta ~
 \eta^{\beta-\alpha-1}(\eta-1)^{\alpha-1}(2sv-\eta m_Q^2)^{n+k-1}e^{-2sv/M_1^2}e^{-2s(1-v)/M_2^2}\Bigg]_{v=1/2}.\nnb\\
\eaeeq

Now, we extend these calculations to the whole terms entering the expressions of the coupling constants under consideration. As the calculations are very lengthy,  as an example, we only present our final result  for the $\frac{3}{2}$ case and the $\tilde{\Pi}_1^{*\mathrm{OPE}}$ function defining  the $\Sigma_b(6097)^+\rightarrow\Lambda_b^0 \pi^+$ transition.  For this function, we get

\begin{eqnarray}\label{magneticmoment2}
\tilde{\Pi}_{1}^{*\mathrm{OPE}}&=&
\int_{m_{b}^{2}}^{s_{0}}e^{-{\frac{s}{M^{2}}}-\frac{m_{\pi}^{2}}{4M^2}}\rho_{1}(s)ds+e^{{-\frac{m_b^2}{M^{2}}}-\frac{m_{\pi}^{2}}{4M^2}}\Gamma_{1},
\end{eqnarray}
where, the expressions, $\rho_{1}(s)$ and $\Gamma_{1}$ are given as:

\begin{eqnarray}
\rho_1(s)&=&\frac{1}{96 \sqrt{2} m_b^2 \pi^2}
\Bigg[-\psi_{31} m_b^4 \mu_{\pi} \zeta_4 + 
    \psi_{31} m_b^4 \mu_{\pi} \beta \zeta_4 + \psi_{20} m_b^4 \mu_{\pi} (\beta-1) (\zeta_4 - 2 \zeta_5) +  2 \psi_{31} m_b^4 \mu_{\pi} \zeta_5 - 2 \psi_{31} m_b^4 \mu_{\pi} \beta \zeta_5 \nonumber\\&+& 12 f_{\pi}\psi_{21} m_{\pi}^2 m_b^3 \zeta_7  + 12 f_{\pi} \psi_{21} m_{\pi}^2 m_b^3 \beta \zeta_7 + 
    12 f_{\pi} \psi_{21} m_{\pi}^2 m_b^2 m_u \beta \zeta_7 - 
    6 f_{\pi} \psi_{21} m_{\pi}^2 m_b^3 \zeta_1 - 6 f_{\pi} \psi_{21} m_{\pi}^2 m_b^2 m_u \zeta_1 \nonumber\\&-& 
    6 f_{\pi} \psi_{21} m_{\pi}^2 m_b^3 \beta \zeta_1 + 3 f_{\pi} \psi{11} m_{\pi}^2 m_b^2 m_u \beta \zeta_1 - 15 f_{\pi} \psi_{12} m_{\pi}^2 m_b^2 m_u \beta \zeta_1 - 12 f_{\pi} \psi_{12} m_{\pi}^2 m_u s \beta \zeta_1 - 12 f_{\pi} \psi_{21} m_{\pi}^2 m_b^3 \zeta_2\nonumber\\& -&  2 f_{\pi} \psi{11} m_{\pi}^2 m_b^2 m_u \zeta_2 + 10 f_{\pi} \psi_{12} m_{\pi}^2 m_b^2 m_u \zeta_2 - 
    8 f_{\pi} \psi_{21} m_{\pi}^2 m_b^2 m_u \zeta_2 + 8 f_{\pi} \psi_{12} m_{\pi}^2 m_u s \zeta_2 - 12 f_{\pi} \psi_{21} m_{\pi}^2 m_b^3 \beta \zeta_2\nonumber\\ 
    &-& 4 f_{\pi}
 \psi_{11} m_{\pi}^2 m_b^2 m_u \beta \zeta_2 + 20 f_{\pi} \psi_{12} m_{\pi}^2 m_b^2 m_u \beta \zeta_2 -  4 f_{\pi} \psi_{21} m_{\pi}^2 m_b^2 m_u \beta \zeta_2 + 16 f_{\pi} 
 \psi_{12} m_{\pi}^2 m_u s \beta \zeta_2 + 8 f_{\pi} \psi_{21} m_{\pi}^2 m_b^2 (2 m_b \nonumber\\
 &+&
  m_u) (2 + \beta) \zeta_6 + 2 \psi_{10} m_b^2 (2 m_b^2 \mu_{\pi} (\zeta_4 + 2 \beta \zeta_4 - (2 + \beta) \zeta_5) +  f_{\pi} m_{\pi}^2 m_u (6 \beta \zeta_7 - 3 \zeta_1 - 4 \zeta_2 - 2 \beta \zeta_2 + 4 (2 + \beta) \zeta_6)\nonumber\\
  & +&  
  f_{\pi} m_{\pi}^2 m_b (6 (\beta + \beta) \zeta_7 - 3 (1 + \beta) (\zeta_1 + 2 \zeta_2) + 8 (2 + \beta) \zeta_6))\Bigg]\Bigg (2 \Big (-6 \beta \zeta_7 + 18 \gamma_E \beta \zeta_7 + 3 \zeta_1 - 9 \gamma_E \zeta_1 - 
 3 \gamma_E \beta \zeta_1 \nonumber\\
 &+& 4 \zeta_2 - 10 \gamma_E \zeta_2 + 2 \beta \zeta_2 -  2 \gamma_E \beta \zeta_2 +  4 (3 \gamma_E -1 ) (2 + \beta) \zeta_6\Big) - \big (18 \beta \zeta_7 - 9 \zeta_1 - 3 \beta \zeta_1 - 10 \zeta_2 - 2 \beta \zeta_2 + 12 (2 \nonumber\\
 &+& \beta) \zeta_6\big)\big (ln(\frac{\Lambda^2}{m_b^2}) +2 ln(\frac{M^2}{\Lambda^2})\big)\Bigg)
+  \frac{1}{32 \sqrt{2} \pi^2}f_{\pi} m_{\pi}^2\Big[  -2 \psi_{10} m_b (1 + \beta) + (\psi_{20} + \psi_{31}) (m_b + m_b \beta - m_u \beta) + 2 m_b (1 \nonumber\\
    &+& \beta)\Bigg] \varphi_{\pi}(u_0)+
 \frac{1}{48 \sqrt{2} \pi^2} (-1 + \tilde{\mu}_{\pi}^2) m_b \mu_{\pi} \Bigg[
   2 (\psi_{10} + \psi_{21}) m_u (\beta-1) + (\psi_{20} + \psi_{31}) m_b \beta\Bigg] \varphi_{\sigma}(u_0) 
\nonumber\\ 
& +&
   \frac{\langle\bar{u}u\rangle}{6 \sqrt{2}}f_{\pi} \beta \varphi_{\pi}(u_0)+
   \frac{\langle g^2G^2\rangle}{288 \sqrt{2} m_b^4 \pi^2}f_{\pi} m_{\pi}^2 m_u (-18 \beta \zeta_{7} + 9 \zeta_1 + 3 \beta \zeta_1 + 10 \zeta_2 + 2 \beta \zeta_2 - 12 (2 + \beta) \zeta_6) \Big[2 - \psi_{01} -\psi_{02} \nonumber\\
   &-& 9 \psi_{10} + 6 \gamma_E \psi_{10} + 3 \psi_{21} + \psi_{22} - 6 \psi_{10} ln(\frac{M^2}{\Lambda^2}) + 
   2 (1 - \psi_{03} + 3 \psi_{21} + 2 \psi_{22} + \psi_{23}) ln(\frac{s-m_b^2}{\Lambda^2})\Big]\nonumber\\
   &+&
   \frac{\langle g^2G^2\rangle}{64 \sqrt{2} m_b^6 \pi^2}
   5 f_{\pi}m_{\pi}^2 m_u \beta \mathbb A(u_0) \Bigg[-m_b^2\Big(6 (2\gamma_E - 3) \psi_{10} + 6 \psi_{21} + 3 \psi_{22} + \psi_{23} - 4 \psi_{-10} + \psi_{-13} + 3 \psi_{-1-1} \nonumber\\
   &-&
   12 \psi_{10} ln(\frac{M^2}{\Lambda^2})\Big) + 
   12  (m_b^2 - s) ln(\frac{s-m_b^2}{\Lambda^2})\Bigg]+
   \frac{\langle g^2G^2\rangle}{576 \sqrt{2} m_b^4 \pi^2}f_{\pi}  \Bigg[(1 - 
      3 (\psi_{10} + \psi_{21})) m_b^3 + \Big(m_b^2 (-3 (\psi_{10} + \psi_{21}) m_b\nonumber\\
      & -& 
         3 (-2 \psi_{01} - 4 \psi_{02} + 3 \psi_{03} - 37 \psi_{10} + 36 \gamma_E \psi_{10} + \psi_{12} + 3 \psi_{13} + 6 \psi_{21} + 2 \psi_{22}) m_u + 
          (m_b + 9 (-3 + 4 \gamma_E) m_u)) \nonumber\\
          &+&
  15  m_u s\Big) \beta + 
   12 m_u \beta \Big(-3 (1 - 3 \psi_{10}) m_b^2 ln(\frac{M^2}{\Lambda^2}) + \big((8 + \psi_{03} - 3 \psi_{21} - 2 \psi_{22} - \psi_{23}) m_b^2 -  6  s\big) ln(\frac{s-m_b^2}{\Lambda^2})\Big)\Bigg] \varphi_{\pi}(u_0)\nonumber\\
   &-&
 \frac{\langle g^2G^2\rangle}{48 \sqrt{2} m_b^3 \pi^2}
  (\tilde{\mu}_{\pi}^2-1) m_u \mu_{\pi} (\beta-1 ) \Bigg[2 - \psi_{01} - \psi_{02} - 9 \psi_{10} + 
    6 \gamma_E \psi_{10} + 3 \psi_{21} + \psi_{22} - 6 \psi_{10} ln(\frac{M^2}{\Lambda^2}) + 
    2 (1 - \psi_{03} \nonumber\\
    &+&
     3 \psi_{21} + 2 \psi_{22} + \psi_{23}) ln(\frac{s-m_b^2}{\Lambda^2})\Bigg] \varphi_{\sigma}(u_0),
\end{eqnarray}

and

\begin{eqnarray}
\Gamma_1&=&-\frac{\langle  \bar{u}u \rangle}{432 \sqrt{2} M^6}
 \Bigg[-12 M^4 \Bigg(m_u M^2 \mu_{\pi} (\zeta_4 + 2 \beta \zeta_4 - (2 + \beta) \zeta_5) + 
 f_{\pi} m_{\pi}^2 m_b m_u (\beta-1) \zeta_2 + f_{\pi} m_{\pi}^2 M^2 (-6 \beta \zeta_7 + 3 \zeta_1 + 4 \zeta_2 \nonumber\\
 &+&
  2 \beta \zeta_2 -4 (2 + \beta) \zeta_6)\Bigg) + 
    m_o^2 m_b \Bigg(2 f_{\pi} m_{\pi}^2 m_b^2 m_u (\beta-1) \zeta_2 - 2 f_{\pi} m_{\pi}^2 m_u M^2 (\beta-1) \zeta_2 + m_b M^2 \big(2 m_u \mu_{\pi} (\zeta_4 + 2 \beta \zeta_4\nonumber\\
    & -&
     (2 + \beta) \zeta_5) + 3 f_{\pi} m_{\pi}^2 (-6 \beta \zeta_7 + 3 \zeta_1 + 4 \zeta_2 + 2 \beta \zeta_2 - 4 (2 + \beta) \zeta_6\big)\Bigg)\Bigg]+
     \frac{\langle  \bar{u}u \rangle}{1728 \sqrt{2} M^8}f_{\pi} m_{\pi}^2 \Bigg[36 M^4 (-2 m_b^2 M^2 \beta - 
      2 M^4 \beta \nonumber\\
      &+&
   m_b^3 m_u (1 + \beta)) + 
   m_o^2 m_b \Big(2m_u M^4 (\beta-1 ) + 18 m_b^3 M^2 \beta - 6 m_b^4 m_u (1 + \beta) + m_b M^4 (5 + 4 \beta) + 
    2 m_b^2 m_u M^2 (7 \nonumber\\
    &+& 5 \beta)\Big)\Bigg] \mathbb A(u_0)-
    \frac{\langle  \bar{u}u \rangle}{432 \sqrt{2} M^4}f_{\pi} \Bigg[36 m_b m_u M^4 (1 + \beta) + 
 m_o^2 \Big (-2 m_b m_u M^2 (\beta- 1 ) + 18 m_b^2 M^2 \beta - 
    6 m_b^3 m_u (1 + \beta) \nonumber\\
    &+& M^4 (5 + 22 \beta)\Big)\Bigg] \varphi_{\pi}(u_0)
   -\frac{\langle  \bar{u}u \rangle}{432 \sqrt{2}
   M^6}(\tilde{\mu}_{\pi}-1 ) (1 + \tilde{\mu}_{\pi}) \mu_{\pi} \Bigg[-12 M^4 \big (-2 m_b M^2 (\beta -1 ) + m_b^2 m_u \beta + m_u M^2 \beta \big) \nonumber\\
   &+&
 m_o^2 m_b^2 \big (-6 m_b M^2 (\beta -1 ) + 2 m_b^2 m_u \beta + 
    m_u M^2 (2\beta -1 )\big)\Bigg] \varphi_{\sigma}(u_0)
    +
    \frac{\langle g^2G^2\rangle}{3456 \sqrt{2} \pi^2}\Bigg[\frac{1}{M^2}
   f_{\pi} m_{\pi}^2 \Big(m_b \big(-6 (1 + \beta) \zeta_7 \nonumber\\
   &+&
    3 (1 + \beta) \zeta_1 + 
 4 (2 + \beta) (\zeta_2 - 2 \zeta_6)\big) + 
      2 m_u (-18 \beta \zeta_7 + 9 \zeta_1 + 3 \beta \zeta_1 + 10 \zeta_2 +
      2 \beta \zeta_2 -12 (2 + \beta) \zeta_6)\Big) \nonumber\\
      &+&
       \frac{1}{
   m_b^2}\Big(-m_b^2 \mu_{\pi} (\zeta_4 + 5 \beta \zeta_4 - 2 (\zeta_5 + 2 \beta \zeta_5)) + 
     2 f_{\pi} m_{\pi}^2 m_b (6 (1 + \beta) \zeta_7 - 3 (1 + \beta) \zeta_1
     -4 (2 + \beta) (\zeta_2 - 2 \zeta_6)) \nonumber\\
     &+& 
     f_{\pi} m_{\pi}^2 m_u (-18 \beta \zeta_7 + 9 \zeta_1 + 3 \beta \zeta_1 + 10 \zeta_2 +2 \beta \zeta_2 - 12 (2 + \beta) \zeta_6)\Big) +
      \frac{1}{M^4}
   f_{\pi} m_{\pi}^2 m_b^2 m_u \Big(2 \big(-6 \beta \zeta_7 +
    18 \gamma_E \beta \zeta_7\nonumber\\
    & +&
     3 \zeta_1 -9 \gamma_E \zeta_1 - 3 \gamma_E \beta \zeta_1 + 4 \zeta_2 -  10 \gamma_E \zeta_2 + 2 \beta \zeta_2 - 2 \gamma_E \beta \zeta_2 + 4 (3 \gamma_E -1 ) (2 + \beta) \zeta_6\big) - \big(18 \beta \zeta_7 - 9 \zeta_1 - 3 \beta \zeta_1\nonumber\\
    & -&
     10 \zeta_2 - 2 \beta \zeta_2 +12 (2 + \beta) \zeta_6\big) \big[ln(\frac{\Lambda^2}{m_b^2}) + 2 ln(\frac{M^2}{\Lambda^2})\big]\Big)\Bigg]+
     \frac{\langle g^2G^2\rangle}{2304 \sqrt{2} m_b^2 M^6 \pi^2} f_{\pi}  m_{\pi}^2 \mathbb A(u_0) \Bigg[
   2 (3 \gamma_E -1 ) m_b^6 m_u \beta \nonumber\\
   &-& 10 m_b^4 m_u M^2 \beta - 
    9 m_b^2 m_u M^4 \beta - 6 m_u M^6 \beta - m_b^3 M^4 (1 + \beta) + 
    2 m_b M^6 (1 + \beta) - 
    3 m_b^6 m_u \beta \big[ln(\frac{\Lambda^2}{mqb^2}) + 2 ln(\frac{M^2}{\Lambda^2})\big]\Bigg]\nonumber\\&+&
     \frac{\langle g^2G^2\rangle}{576 \sqrt{2}  M^2 \pi^2}
    f_{\pi}  m_u \beta \Bigg[(2 - 6 \gamma_E) m_b^2 - 
    6 (\gamma_E- 1) M^2 + 3 (m_b^2 + 2 M^2) ln(\frac{\Lambda^2}{m_b^2}) + 
    3 (2 m_b^2 + 3 M^2) \ln(\frac{M^2}{\Lambda^2})\Bigg] \varphi_{\pi}(u_0)\nonumber\\
    &+&
    \frac{\langle g^2G^2\rangle}{1728 \sqrt{2} m_b M^4 \pi^2} (\tilde{\mu}_{\pi}^2 -1 ) \mu_{\pi} \Bigg[
   2 (3 \gamma_E- 1 ) m_b^4 m_u (\beta- 1) -  7 m_b^2 m_u M^2 (\beta -1) - m_u M^4 (\beta -1 ) - m_b M^4 \beta \nonumber\\ &-&
    3 m_b^4 m_u (\beta- 1 ) \Big[
       ln[\frac{\Lambda^2}{m_b^2}) + 2 ln(\frac{M^2}{\Lambda^2})\Big]\Bigg] \varphi_{\sigma}(u_0)+
       \frac{\langle g^2G^2\rangle \langle  \bar{u}u \rangle}{20736 \sqrt{2}
   M^{14}} f_{\pi} m_{\pi}^2 m_b \Bigg[-6 M^4 \Big (-2 m_b^3 M^2 \beta + 
  4 m_b M^4 \beta \nonumber\\ &+& m_b^4 m_u (1 
  + \beta) - 6 m_b^2 m_u M^2 (1 + \beta) + 
       6 m_u M^4 (1 + \beta)\Big) + 
    m_o^2 \Big (-3 m_b^5 M^2 \beta + 18 m_b^3 M^4 \beta - 18 m_b M^6 \beta + 
        m_b^6 m_u (1 \nonumber\\
        &+& \beta) - 11 m_b^4 m_u M^2 (1 + \beta) + 
        30 m_b^2 m_u M^4 (1 + \beta) - 
        18 m_u M^6 (1 + \beta)\Big)\Bigg] \mathbb A(u_0)
    -\frac{\langle g^2G^2\rangle \langle  \bar{u}u \rangle}{5184 \sqrt{2} M^{10}} f_ {\pi}  m_b \Bigg[
   6 M^4 (2 m_b M^2 \beta \nonumber\\
   &-& m_b^2 m_u (1 + \beta) + 3 m_u M^2 (1 + \beta)) + 
    m_o^2 \big(-3 m_b^3 M^2 \beta + 6 m_b M^4 \beta + m_b^4 m_u (1 + \beta) - 
        6 m_b^2 m_u M^2 (1 + \beta) \nonumber\\
        &+& 
        6 m_u M^4 (1 + \beta)\big)\Bigg] \varphi_{\pi}(u_0)+        
    \frac{\langle g^2G^2\rangle \langle  \bar{u}u \rangle}{15552 \sqrt{2}
   M^{12}}  (\tilde{\mu}_{\pi}^2 -1 ) m_b \mu_{\pi} \Bigg[
   m_o^2 (m_b^4 - 6 m_b^2 M^2 + 6 M^4) (-3 M^2 (\beta -1) + 
       m_b m_u \beta) \nonumber\\
       &-& 
    6 M^4 \big (2 (m_b^2 - 3 M^2) M^2 + (m_b^3 m_u - 
           2 m_b (m_b + m_u) M^2 + 6 M^4) \beta \big)\Bigg] \varphi_{\sigma}(
  u_0).
\end{eqnarray}

In the above functions $\zeta_{j}$ and $\psi_{nm}$ are defined as
\begin{eqnarray}\label{etalar}
\zeta_{j} &=& \int {\cal D}\alpha_i \int_0^1 dv f_{j}(\alpha_i)
\delta(k(\alpha_{ q} +v \alpha_g) -  u_0),
\nonumber \\
\nonumber \\
\psi_{nm}&=&\frac{{( {s-m_{Q}}^2 )
}^n}{s^m{(m_{Q}^{2})}^{n-m}},\nonumber \\
\end{eqnarray}
 with the  distribution amplitude given as $f_{1}(\alpha_i)={\cal V_{\parallel}}(\alpha_i)$, $f_{2}(\alpha_i)={\cal V_{\perp}}(\alpha_i)$,
 $f_{3}(\alpha_i)={\cal A_{\parallel}}(\alpha_i)$, $f_{4}(\alpha_i)={\cal
 T}(\alpha_i)$, $f_{5}(\alpha_i)=v{\cal T}(\alpha_i)$, $f_{6}(\alpha_i)=v{\cal V_{\perp}}(\alpha_i)$, $f_{7}(\alpha_i)=v{\cal A_{\parallel}}(\alpha_i)$,  whose explicit forms can be found in Refs.~\cite{Belyaev:1994zk,Ball:2004ye,Ball:2004hn}. As we previously mentioned,  $u_{0}$ has the form, $u_{0}=\frac{M_{1}^{2}}{M_{1}^{2}+M_{2}^{2}}$. Considering the close masses of initial and final baryons and taking $ M_{1}^{2} = M_{2}^{2} $, it  becomes,
 $u_{0} =\frac{1}{2}$. In the above results, $\mu_{\pi} = f_{\pi} \frac{m_{\pi}^2}{m_{u} + m_{d}},$ $\tilde
\mu_{\pi} = \frac{{m_{u} + m_{d}}}{m_{\pi}}$ and ${\cal D} \alpha =
 d \alpha_{\bar q}  d \alpha_q  d \alpha_g
\delta(1-\alpha_{\bar q}-\alpha_q-\alpha_g)$ are used. The functions 
$\varphi_{\pi}(u),$ $\mathbb A(u),$ $\mathbb B(u),$ $\varphi_P(u),$
$\varphi_\sigma(u),$ ${\cal T}(\alpha_i),$ ${\cal
A}_\perp(\alpha_i),$ ${\cal A}_\parallel(\alpha_i),$ ${\cal
V}_\perp(\alpha_i)$ and ${\cal V}_\parallel(\alpha_i)$ are functions
of definite twists, which can also be found in Refs~ \cite{Belyaev:1994zk,Ball:2004ye,Ball:2004hn}. They are given as
\begin{eqnarray}
\varphi_{\pi}(u) &=& 6 u \bar u \left( 1 + a_1^{\pi} C_1(2 u -1) +
a_2^{\pi} C_2^{3 \over 2}(2 u - 1) \right),
\nonumber \\
{\cal T}(\alpha_i) &=& 360 \eta_3 \alpha_{\bar q} \alpha_q
\alpha_g^2 \left( 1 + w_3 \frac12 (7 \alpha_g-3) \right),
\nonumber \\
\varphi_P(u) &=& 1 + \left( 30 \eta_3 - \frac{5}{2} \mu_{\pi}^2 \right)
C_2^{1 \over 2}(2 u - 1)
\nonumber \\
&+& \left( -3 \eta_3 w_3  - \frac{27}{20} \mu_{\pi}^2 -
\frac{81}{10} \mu_{\pi}^2 a_2^{\pi} \right) C_4^{1\over2}(2u-1),
\nonumber \\
\varphi_\sigma(u) &=& 6 u \bar u \left[ 1 + \left(5 \eta_3 - \frac12
\eta_3 w_3 - \frac{7}{20}  \mu_{\pi}^2 - \frac{3}{5} \mu_{\pi}^2
a_2^{\pi} \right) C_2^{3\over2}(2u-1) \right],
\nonumber \\
{\cal V}_\parallel(\alpha_i) &=& 120 \alpha_q \alpha_{\bar q}
\alpha_g \left( v_{00} + v_{10} (3 \alpha_g -1) \right),
\nonumber \\
{\cal A}_\parallel(\alpha_i) &=& 120 \alpha_q \alpha_{\bar q}
\alpha_g \left( 0 + a_{10} (\alpha_q - \alpha_{\bar q}) \right),
\nonumber\\
{\cal V}_\perp (\alpha_i) &=& - 30 \alpha_g^2\left[
h_{00}(1-\alpha_g) + h_{01} (\alpha_g(1-\alpha_g)- 6 \alpha_q
\alpha_{\bar q}) +
    h_{10}(\alpha_g(1-\alpha_g) - \frac32 (\alpha_{\bar q}^2+ \alpha_q^2)) \right],
\nonumber\\
{\cal A}_\perp (\alpha_i) &=& 30 \alpha_g^2(\alpha_{\bar q} -
\alpha_q) \left[ h_{00} + h_{01} \alpha_g + \frac12 h_{10}(5
\alpha_g-3) \right],
\nonumber \\
\mathbb B(u)&=& g_{\pi}(u) - \phi_{\pi}(u),
\nonumber \\
g_{\pi}(u) &=& g_0 C_0^{\frac12}(2 u - 1) + g_2 C_2^{\frac12}(2 u -
1) + g_4 C_4^{\frac12}(2 u - 1),
\nonumber \\
{\mathbb A}(u) &=& 6 u \bar u \left[\frac{16}{15} + \frac{24}{35}
a_2^{\pi}+ 20 \eta_3 + \frac{20}{9} \eta_4 +
    \left( - \frac{1}{15}+ \frac{1}{16}- \frac{7}{27}\eta_3 w_3 - \frac{10}{27} \eta_4 \right) C_2^{3 \over 2}(2 u - 1)
    \right. \nonumber \\
    &+& \left. \left( - \frac{11}{210}a_2^{\pi} - \frac{4}{135} \eta_3w_3 \right)C_4^{3 \over 2}(2 u - 1)\right]
\nonumber \\
&+& \left( -\frac{18}{5} a_2^{\pi} + 21 \eta_4 w_4 \right)\left[ 2
u^3 (10 - 15 u + 6 u^2) \ln u \right. \nonumber\\ &+& \left. 2 \bar
u^3 (10 - 15 \bar u + 6 \bar u ^2) \ln\bar u + u \bar u (2 + 13 u
\bar u) \right] \label{wavefns},
\end{eqnarray}
where $C_n^k(x)$ are the Gegenbauer polynomials,
\begin{eqnarray}
h_{00}&=& v_{00} = - \frac13\eta_4,
\nonumber \\
h_{01} &=& \frac74  \eta_4 w_4  - \frac{3}{20} a_2^{\pi},
\nonumber \\
h_{10} &=& \frac74 \eta_4 w_4 + \frac{3}{20} a_2^{\pi},
\nonumber \\
a_{10} &=& \frac{21}{8} \eta_4 w_4 - \frac{9}{20} a_2^{\pi},
\nonumber \\
v_{10} &=& \frac{21}{8} \eta_4 w_4,
\nonumber \\
g_0 &=& 1,
\nonumber \\
g_2 &=& 1 + \frac{18}{7} a_2^{\pi} + 60 \eta_3  + \frac{20}{3}
\eta_4,
\nonumber \\
g_4 &=&  - \frac{9}{28} a_2^{\pi} - 6 \eta_3 w_3 \label{param0}.
\end{eqnarray}
The constants presented  in Eqs.~(\ref{wavefns}) and (\ref{param0}) are
calculated using QCD sum rules at the renormalization scale $\mu=1$~GeV$^{2}$ \cite{Ball:2006wn,Belyaev:1994zk,Ball:2004ye,Ball:2004hn,R23,ek1,ek2,ek3}. These constants are given as  $a_{1}^{\pi} = 0$,
$a_{2}^{\pi} = 0.44$, $\eta_{3} =0.015$, $\eta_{4}=10$, $w_{3} = -3$
and $ w_{4}= 0.2$.

\section*{ACKNOWLEDGMENTS}

H. S. thanks Kocaeli University for the partial financial support through the grant BAP 2018/070.

\label{sec:Num}
%%%%%%%%%%%%%%%%%%%%%%%%%%%%%%%%%%%%%%%%%%%%%%%%%%%%%%%%%%%%%%%%%%%%%%%%%%%%%%%%%%%%%%%%%%%%%%%%%%%%%%%%%%%%

\end{document}